\pdfoutput=1

\documentclass[11pt]{article}

\usepackage[final]{acl}

\usepackage{times}
\usepackage{latexsym}

\usepackage[T1]{fontenc}

\usepackage[utf8]{inputenc}

\usepackage{microtype}

\usepackage{inconsolata}

\usepackage[table]{xcolor}
\usepackage{tikz}
\usepackage{nicematrix}

\usepackage[utf8]{inputenc}
\usepackage{booktabs,multirow}
\usepackage{enumitem}
\usepackage{kotex}
\usepackage{color}
\usepackage{rotating}
\usepackage{amsmath}
\usepackage{amsfonts}
\usepackage{pifont}
\usepackage{array}
\usepackage{amssymb}
\usepackage{algpseudocode}
\usepackage{algorithm}
\usepackage{float}
\usepackage{tcolorbox}
\usepackage{multicol}
\usepackage{lipsum}
\usepackage[normalem]{ulem}
\usepackage{xcolor}
\usepackage{soul}
\usepackage{listings}
\usepackage{amsmath}
\usepackage{adjustbox}
\usepackage{graphicx} 
\usepackage{colortbl} 
\usepackage[dvipsnames]{xcolor}   
\usepackage{mdframed}

\definecolor{LightCyan}{rgb}{0.88,1,1}

\definecolor{customhighlight}{HTML}{DDB6E4}
\sethlcolor{customhighlight}

\colorlet{instructionbg}{gray!15}
\colorlet{questionbg}{gray!25}

\algrenewcommand\algorithmicrequire{\textbf{Input:}}
\algrenewcommand\algorithmicensure{\textbf{Output:}}

\newcommand{\cmark}{\ding{51}}
\newcommand{\xmark}{\ding{55}}

\newcommand{\ie}{{\it i.e.}}
\newcommand{\eg}{{\it e.g.}}

\newcommand{\ours}{G{\small RUT}}

\newcommand{\valfont}[1]{{\fontsize{6}{7}\selectfont #1}}
\newcommand{\cellimg}[1]{%
  \includegraphics[height=0.85cm, keepaspectratio]{#1}%
}

\title{Enhancing Time Awareness in Generative Recommendation}

\author{Sunkyung Lee, Seongmin Park, Jonghyo Kim, Mincheol Yoon, Jongwuk Lee\thanks{\ \ Corresponding author} \\
        Sungkyunkwan University, Republic of Korea\\  
        \texttt{ \{sk1027, psm1206, naye971012, yoon56, jongwuklee\}@skku.edu
        }}

\begin{document}
\maketitle
\begin{abstract}
Generative recommendation has emerged as a promising paradigm that formulates the recommendations into a text-to-text generation task, harnessing the vast knowledge of large language models. However, existing studies focus on considering the sequential order of items and neglect to handle the \emph{temporal dynamics} across items, which can imply evolving user preferences. To address this limitation, we propose a novel model, \emph{\textbf{G}enerative \textbf{R}ecommender \textbf{U}sing \textbf{T}ime awareness (\textbf{\ours})}, effectively capturing hidden user preferences via various temporal signals. We first introduce \textit{Time-aware Prompting}, consisting of two key contexts. The user-level temporal context models personalized temporal patterns across timestamps and time intervals, while the item-level transition context provides transition patterns across users. We also devise \textit{Trend-aware Inference}, a training-free method that enhances rankings by incorporating trend information about items with generation likelihood. Extensive experiments demonstrate that \ours~outperforms state-of-the-art models, with gains of up to 15.4\% and 14.3\% in Recall@5 and NDCG@5 across four benchmark datasets. The source code is available at \url{https://github.com/skleee/GRUT}.

\end{abstract}

\section{Introduction}\label{sec:introduction}

Generative recommendation (GR) is an emerging paradigm that redefines the traditional recommendation task as a text-to-text generation problem~\cite{RajputMSKVHHT0S23TIGER, Geng0FGZ22P5}. While the conventional discriminative approach ranks items individually~\cite{KangM18SASRec}, GR directly generates the identifier (ID) of the target item given a user history. Notably, it benefits from directly leveraging the extensive capabilities of large language models (LLMs) for recommendations~\cite{JMLR/Raffel2020/T5, TouvronLIMLLRGHARJGL23LLaMA}.

\begin{figure}[t]
\centering
\includegraphics[width=1.0\linewidth]{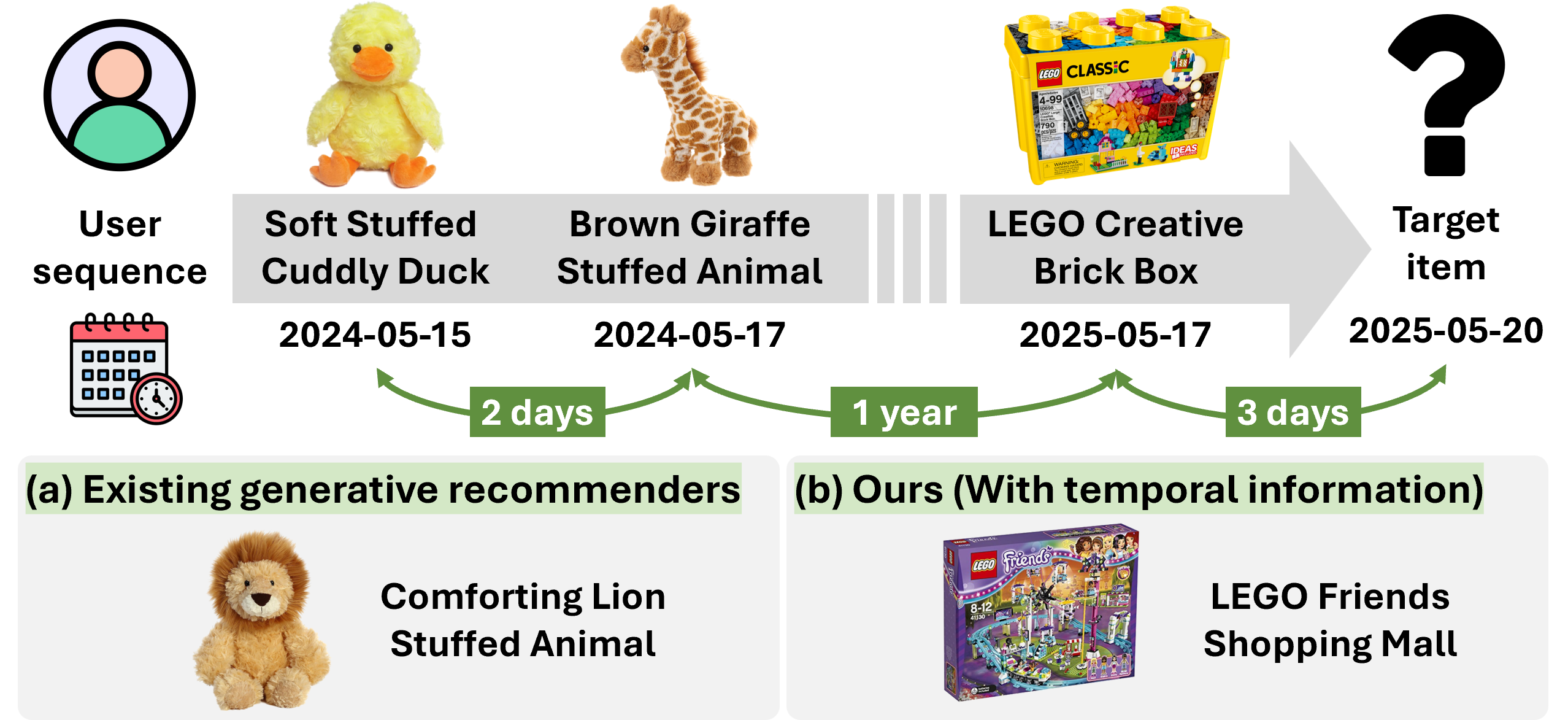}
\vspace{-3.5mm}
\caption{Illustration of our motivation. While (a) existing generative recommenders only consider sequential order, (b) our method utilizes temporal information.}\label{fig:motivation}
\vspace{-3.5mm}
\end{figure}

Despite its success, existing GR models overlook a crucial dimension: \textit{temporal dynamics}. As illustrated in Figure~\ref{fig:motivation}, the temporal information of items significantly affects user preferences~\cite{LiWM20TISASREC, aaai/Zhang24TGCL4SR}. Without temporal information, the model might recommend another `\textit{stuffed animal}' based on frequent occurrences in the user history, even after preference has shifted toward the `\textit{LEGO product}' over one year (Figure~\ref{fig:motivation}a). In contrast, a time-aware GR model can accurately discern preference shifts and recommend products that match the user's current interests by considering temporal dynamics (Figure~\ref{fig:motivation}b). Moreover, timestamps may imply seasonal preferences that the mere item order cannot capture, \eg, Christmas or holidays~\cite{wsdm/Wang20OAR}.

Incorporating temporal information into recommendations yields several challenges. (i) Temporal signals exist in distinct forms: \emph{absolute timestamps} and \emph{relative time intervals} across user interactions. Each provides different signals, making it challenging to preserve their information while effectively combining them~\cite{recsys/Cho20MEANTIME, www/Zhang25HM4SR}. (ii) Temporal item patterns vary in scope from individual user behavior to collective item-level trends and transition patterns. The collective patterns further require analyzing the interaction of all users. While previous work has adopted graph-based methods~\cite{aaai/Zhang24TGCL4SR, sigir/Wang20GCEGNN}, representing temporal knowledge in natural language form for GR remains unexplored. (iii) Integrating temporal signals into GR requires unique modeling. Unlike traditional sequential models that rely on explicit temporal embeddings~\cite{LiWM20TISASREC, www/Zhang25HM4SR, TOIS/Hu25HORAE} or contrastive learning~\cite{cikm/Tian22TCPSRec, aaai/Dang23TiCoSeRec, aaai/Zhang24TGCL4SR}, it is crucial to translate complex temporal patterns into natural language. Concurrently, GR models are required to maintain the ability to generate precise item IDs from the vast item candidate pool. 

To address these challenges, we propose a novel model, \emph{\textbf{G}enerative \textbf{R}ecommender \textbf{U}sing \textbf{T}ime awareness (\textbf{\ours})}, enhancing GR through temporal signals of items. To model distinct temporal signals, we first introduce \textit{Time-aware Prompting}, which consists of two contexts. At the user level, we integrate absolute timestamps and time intervals between interactions in the prompt to model individual user patterns. At the item level, item transition patterns are represented in natural language forms, incorporating broader temporal patterns that individual user history alone cannot provide. Besides, we devise \textit{Trend-aware Inference}, a flexible method that refines beam search ranking with the temporal trend of items. It adaptively combines item generation likelihoods with trend scores, assigning higher scores to recently trending items. Despite its simplicity, it enables the model to reflect diverse and timely recommendation scenarios. 

Our key contributions are summarized as follows: (i) To the best of our knowledge, this is the first work to integrate temporal dynamics into GR, demonstrating its importance beyond the mere sequential order of items. (ii) We propose \textit{Time-aware Prompting}, which effectively incorporates multi-dimensional temporal patterns at both user and item levels. (iii) We design \textit{Trend-aware Inference}, which adaptively leverages trends to refine recommendation rankings without model retraining. (iv) Extensive experiments demonstrate that \ours~outperforms state-of-the-art models, with improvements of up to 15.4\% in Recall@5 and 14.3\% in NDCG@5 across four benchmark datasets.

\section{Related Work}

\begin{table}[]\scriptsize
\centering
\setlength{\tabcolsep}{2.8pt}
\renewcommand{\arraystretch}{0.85}
\begin{footnotesize}
\begin{tabular}{c|c|c}
\toprule
 & Discriminative & Generative \\ \midrule
\begin{tabular}[c]{@{}c@{}}Sequential \\ info. \end{tabular} & 

\begin{tabular}[c]{@{}c@{}}
GRU4Rec, HGN, \\ 
SASRec, BERT4Rec,  \\ 
FDSA, S$^3$Rec  \\
\end{tabular} & 

\begin{tabular}[c]{@{}c@{}} 
P5, TIGER, \\ 
LC-Rec, LETTER, \\
IDGenRec, TransRec, \\
ELMRec, GRAM \\

\end{tabular} \\ \midrule

\begin{tabular}[c]{@{}c@{}}Temporal\\ info. \end{tabular}  & 

\begin{tabular}[c]{@{}c@{}}
TiSASRec, TGSRec, \\
TCPSRec, TiCoSeRec, \\
TGCL4SR, \\
HM4SR, HORAE  \\
\end{tabular} & 

\textbf{G{\scriptsize RUT} (Ours)} \\ 
\bottomrule
\end{tabular}
\end{footnotesize}
\caption{Category of existing sequential recommendation models. \ours~introduces a time-aware generative recommendation model.}\label{tab:category}
\vspace{-3.5mm}

\end{table}

We categorize sequential recommendation\footnote{For more details on existing sequential recommendation models, see Appendix~\ref{sec:app_relatedwork_sr}.} into two dimensions, as shown in Table~\ref{tab:category}: temporal information utilization and whether they employ generative approaches~\cite{SurveyGenSearchRecSys}.

\subsection{Generative Recommendation}

In line with recent advancements in generative search~\cite{nips/Tay/DSI, emnlp23GLEN}, generative recommendation has emerged as a paradigm that directly generates the target item identifier from user history\footnote{We mainly focus on improving the GR model, aiming to generate target item identifiers. See Appendix~\ref{sec:app_relatedwork_llmrec} for further LLM-based recommendation models.}. 
P5~\cite{Geng0FGZ22P5, HuaXGZ23P5Howtoindex} first pioneered this paradigm with multi-task learning. Recent works have largely focused on item identifiers. TIGER~\cite{RajputMSKVHHT0S23TIGER}, LC-Rec~\cite{ZhengHLCZCW24LCRec}, and LETTER~\cite{cikm24LETTER} use vector quantization~\cite{ZeghidourLOST22RQVAE} for codebook-based identifiers. LC-Rec further aligns language and collaborative semantics with codebook IDs, and LETTER integrates collaborative signals into identifiers. Meanwhile, IDGenRec~\cite{Tan24IDGenRec} generates keyword IDs from textual metadata, and TransRec~\cite{kdd24TransRec} combines multiple identifier types. ELMRec~\cite{WangCFS24ELMRec} injects graph-based high-order interaction knowledge. More recently, GRAM~\cite{acl25GRAM} translates item relationships using textual identifiers and employs late fusion to integrate item semantics. However, the temporal dynamics of items remain unexplored in GR, which struggles to grasp shifting user preferences over time.

\begin{figure*}[t]
\centering
\includegraphics[width=1.0\linewidth]{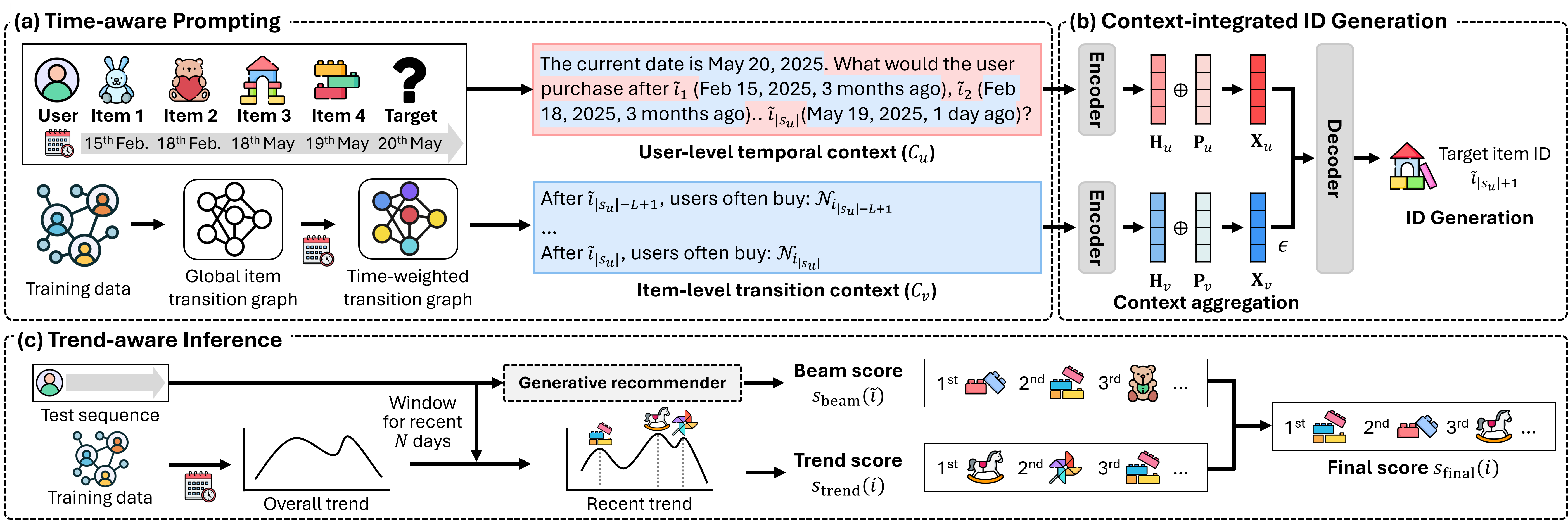}
\vspace{-5.5mm}
\caption{Overall architecture of \ours. The core innovation of the model is  (a) \textit{Time-aware Prompting} that captures evolving user preferences. This is followed by (b) \textit{Context-integrated ID Generation} that aggregates contexts and complemented by (c) \textit{Trend-aware Inference} that adaptively incorporates the trend of items.}\label{fig:overall_architecture}
\vspace{-3.5mm}
\end{figure*}

\subsection{Temporal Recommendation}
Temporal information in recommendations implies how user preferences evolve, providing richer information than the sequential order of items in the sequence. TiSASRec~\cite{LiWM20TISASREC} initiated the use of time intervals with self-attention~\cite{KangM18SASRec}, while TGSRec~\cite{cikm/Fan21TGSRec} incorporates timestamp embeddings. 
Several models leverage contrastive learning with temporal information: TCPSRec~\cite{cikm/Tian22TCPSRec} employs temporal contrastive pre-training, TiCoSeRec~\cite{aaai/Dang23TiCoSeRec} develops time-aware sequence augmentation methods, and TGCL4SR~\cite{aaai/Zhang24TGCL4SR} constructs temporal item transition graphs for graph-based contrastive learning. Recent work extends temporal dynamics to non-neural models~\cite{wsdm/Park25TALE}.

Importantly, for multi-modal sequential recommendation, HM4SR~\cite{www/Zhang25HM4SR} encodes timestamps into embeddings and combines them with item ID, text, and image representations through a mixture of experts. HORAE~\cite{TOIS/Hu25HORAE} enhances a multi-interest pre-training model by incorporating temporal context with texts. However, both models only extract representations from LLMs without fine-tuning, limiting their ability to fully harness the capabilities of LLMs. Recent work has also explored temporal awareness for LLMs in sequential recommendation~\cite{corr/24Tempura}. However, it only evaluates on sampled candidates rather than the entire set, limiting its scalability.

\section{Background} \label{sec:background}

Let $\mathcal{U}$ and $\mathcal{I}$ denote the sets of users and items, respectively. Each user $u \in \mathcal{U}$ has an interaction history represented as a sequence $s_u=(i_1, \dots, i_{|s_u|})$, where each interaction corresponds to actions such as purchasing or clicking. Each element $i_j$ represents the item the user interacted with at the $j$-th position, and $|s_u|$ indicates the number of items in $s_u$. The timestamp sequence is denoted by $T_u=(t_1, \dots, t_{|s_u|})$, indicating the temporal information corresponding to $s_u$. Sequential recommendation aims to predict the next item $i_{|s_u|+1} \in \mathcal{I}$ that the user is most likely to interact with.

For generative recommendation, each item $i \in \mathcal{I}$ is assigned a unique ID $\tilde{i}$. Generally, item IDs can be represented as a sequence of codebook tokens~\cite{RajputMSKVHHT0S23TIGER, ZhengHLCZCW24LCRec} or short text~\cite{Tan24IDGenRec}. With the item ID, the user sequence is converted to the sequence of item IDs $\tilde{s}_u=(\tilde{i}_1, \tilde{i}_2, \dots, \tilde{i}_{|s_u|})$. Following~\citet{acl25GRAM}, we extract keywords from item descriptions using term frequency~\cite{Jones2004/TFIDF} to create descriptive item IDs.\footnote{See Appendix~\ref{sec:app_expsetup_detail_grut} for details of keyword extraction.} In existing studies~\cite{Geng0FGZ22P5, Tan24IDGenRec}, the user sequence is represented without temporal information:
\begin{tcolorbox}
What would the user purchase after $\tilde{i}_1$, $\tilde{i}_2$, \dots, $\tilde{i}_{|s_u|}$?
\end{tcolorbox}

\noindent
Here, the goal is to generate the target item ID $\tilde{i}_{|s_u|+1}$, which the user is most likely to prefer.

\section{Proposed Model} \label{sec:method}

We present a \emph{\textbf{G}enerative \textbf{R}ecommender \textbf{U}sing \textbf{T}ime awareness (\textbf{\ours})}, which enhances GR via explicit modeling of temporal dynamics. The overall architecture is depicted in Figure~\ref{fig:overall_architecture}. Our primary contribution is \textit{Time-aware Prompting} that effectively captures temporal patterns from both individual user behavior and collective user transitions (Section~\ref{sec:time_prompting}). These patterns are then utilized in \textit{Context-integrated ID Generation} (Section~\ref{sec:id_generation}). After training, we design \textit{Trend-aware Inference}, which refines rankings by incorporating generation likelihood with temporal trends (Section~\ref{sec:trend_inference}).

\subsection{Time-aware Prompting}\label{sec:time_prompting}

We introduce time-aware prompting that harnesses temporal dynamics by incorporating user-level temporal context and item-level transition patterns. It models individual temporal patterns based on absolute timestamps and relative intervals while leveraging collective transition patterns across users. 

\subsubsection{User-level Temporal Context}\label{sec:local_prompt}
The user-specific temporal patterns are encoded into natural language form, leveraging the capabilities of LLMs to process temporal information through prompting~\cite{acl/Xiong24TGLLM}. We enhance the basic input of GR by injecting the temporal information of interactions.

Specifically, we utilize two distinct forms of temporal signals: \textit{target-relative intervals} and \textit{absolute timestamps}. (i) The target-relative interval $\Delta t_i$ (\ie, the interval between each timestamp $t_i$ and the inference timestamp $t_{|s_u|+1}$) effectively reflects how user preferences may have shifted over time. For instance, recent interests can be highlighted when recommending shortly after an interaction, while stable long-term preferences are emphasized for longer intervals. (ii) Absolute timestamps $t_i$ enable the model to recognize seasonal patterns or cyclical behaviors, \eg, holiday shopping, that intervals alone cannot capture.

The user-level temporal context $C_u$ is as follows:
\begin{tcolorbox}
The current date is $t_{|s_u|+1}$. \\ What would the user purchase after \\
$\tilde{i}_1$ ($t_1$, $\Delta t_1$ ago), 
$\tilde{i}_2$ ($t_2$, $\Delta t_2$ ago), 
$\cdots$, \\
$\tilde{i}_{|s_u|}$ ($t_{|s_u|}$, $\Delta t_{|s_u|}$ ago) ?
\end{tcolorbox}

\noindent
where $\tilde{i}$ represents item IDs that can take various forms, \eg, keywords or titles. Note that our method can be generally applied regardless of the item ID representations, as shown in Appendix~\ref{sec:app_expresult_id}.

This context enables learning complex patterns beyond the sequential orders of items, grasping preference shifts according to time intervals. Notably, it has three key advantages: (i) The target-relative intervals and absolute timestamps provide complementary signals that consistently outperform either when used alone, as shown in Table~\ref{tab:exp_time_info}. (ii) Owing to the target-relative interval, it is particularly effective for long time intervals, where user preferences have likely evolved, as demonstrated in Figure~\ref{fig:exp_time_interval_group}. (iii) By considering current dates, recommendations can be dynamically adapted based on inference timestamps, unlike existing GR models that make identical predictions regardless of inference timestamps. It is reported in Appendix~\ref{sec:app_case_shift}.

\subsubsection{Item-level Transition Context}\label{sec:item_prompt}
We leverage item-level transition patterns to capture common consumption behaviors across all users, identifying what items users typically consume next after specific items. While user-level temporal context focuses on individual preferences over time, it cannot model collective patterns across users. The item transition pattern has been widely recognized as crucial information in recommendations~\cite{aaai/Zhang24TGCL4SR, sigir/Wang20GCEGNN}. Apart from previous studies, we convert these structural patterns into natural language formats for GR.

\noindent
\textbf{Global Item Transition Graph}. 
We first construct a global item transition graph $\mathcal{G}=(\mathcal{V}, \mathcal{E})$ from all training sequences. Here, the node set $\mathcal{V}$ represents all items, and the edge set $\mathcal{E}$ represents transitions between items. For each user sequence $s_u$, we extract all item pairs $(i_t, i_{t'})$ where $t<t'$ and record the time interval $\Delta t_{i,j} = t_j - t_i$. We add all pairs as directed edges to the graph, where each edge $e_{i,j} \in \mathcal{E}$ denotes a transition from item $i$ to item $j$, along with the corresponding time interval $\Delta t_{i,j}$. 

\noindent
\textbf{Time-weighted Transition Graph}. 
For a given item $i \in \mathcal{I}$, we calculate transition probabilities for all outgoing edges $ \{ e_{i,j}|j\in \mathcal{I} \}$ from the graph, considering time intervals~\cite{wsdm/Park25TALE}. We assign a time-decaying weight that gives higher importance to shorter time intervals:
\begin{equation}\label{eq:time_weight}
w(\Delta t_{i,j}) = \max\left(\exp\left(-\frac{|\Delta t_{i,j}|}{\tau}\right), c\right),
\end{equation}

\noindent
where $\tau$ controls decay speed and $c$ ensures minimum weight for long-term transitions. Using time-aware weights, the transition probability $p_{i,j}$ is formulated as:
\begin{equation}\label{eq:global_count}
p_{i,j} = \frac{ \sum_{(i,j)\in\mathcal{A}_{i,j}} w(\Delta t_{i,j}) }{\sum_{j' \in \mathcal{I}} \sum_{(i,j')\in\mathcal{A}_{i,j'}}  w( \Delta t_{i,j'})},
\end{equation}
\noindent
where $\mathcal{A}_{i,j}$ denotes the set of all pairs from item $i$ to item $j$ in the training data. 

Based on the transition probability, we extract meaningful patterns by selecting the top-$k$ neighboring items for each of the last $L$ items in the sequence:
\begin{equation}\label{eq:global_topk}
\mathcal{N}_i = \{\tilde{i}^1, ..., \tilde{i}^k\} = \mathop{\text{Top-}k}_{j \in \mathcal{I}}p_{i,j},
\end{equation}
\noindent
where $\mathcal{N}_i$ represents the set of top-$k$ neighboring item IDs for item $i$. Here, $k$ and $L$ are hyperparameters. We focus on the last $L$ items in the sequence, considering the recency and the maximum input sequence length of language models. These top-$k$ items refer to the items that users most frequently purchased after the given item, based on the collective behavior patterns across all users.

\noindent
\textbf{Prompt Transformation}.
The extracted transition patterns are then transformed into natural language using item IDs. The item-level transition context $C_v$ is expressed as:
\begin{tcolorbox}[left=10pt, right=10pt]
After $\tilde{i}_{|s_u|-L+1}$, users often buy: $\mathcal{N}_{i_{|s_u|-L+1}}$. \\
$\cdots$ \\
After $\tilde{i}_{|s_u|}$, users often buy: $\mathcal{N}_{i_{|s_u|}}$. 
\end{tcolorbox}
\noindent
where the item $\tilde{i}_{|s_u|-L+n}$ is the $n$-th item among the last $L$ items, and $\mathcal{N}_{i_{|s_u|-L+n}}$ is represented by concatenating all item IDs within the set.
This context can integrate the item transition knowledge with the recommendation process, in addition to the user-specific temporal context.

\subsection{Context-integrated ID Generation}\label{sec:id_generation}

After extracting user-level temporal patterns and item-level transition knowledge, we aggregate the two contexts to generate accurate target item IDs that reflect evolving user preferences.

\noindent
\textbf{Context Aggregation}.
We employ a well-established parallel encoding approach~\cite{eacl/IzacardG21/FiD, acl/Yen24CEPE, acl25GRAM}. It consists of two key steps: (i) encoding each context independently and (ii) aggregating contexts in the decoder through cross-attention.
First, the user-level temporal context $C_u$ and the item-level transition context $C_v$ are separately encoded with a shared encoder:
\begin{equation}\label{eq:encoder_output}
    \mathbf{H}_c=\text{Encoder}\left(C_c\right) \in \mathbb{R}^{M \times d}, \quad c \in \{u, v\}
\end{equation}
\noindent
where $M$ represents the number of text tokens, and $d$ is the hidden dimension size. To further distinguish context types, learnable context-type embeddings are added to encoder outputs:
\begin{equation}\label{eq:context_emb}
\mathbf{X}_c = \mathbf{H}_c + \mathbf{P}_c \in \mathbb{R}^{M \times d}, \quad c \in \{u, v\}
\end{equation}
where $\mathbf{P}_u$ and $\mathbf{P}_v$ are unique embeddings for user-level and item-level contexts, respectively.
We then combine all representations into a unified embedding matrix $\mathbf{X}$:
\begin{equation}\label{eq:key_value_matrix}
    \mathbf{X} = [ \mathbf{X}_u; \epsilon \cdot \mathbf{X}_v ] \in \mathbb{R}^{( 2 \times M) \times d},
\end{equation}
where $\epsilon$ is a hyperparameter that controls the effect of transition patterns without overwhelming user-specific signals. 
Finally, the decoder processes the unified information via cross-attention, where $\mathbf{X}$ serves as the key-value matrix.

\noindent
\textbf{Training Objective}.
Once contexts are aggregated, the decoder autoregressively generates the target item ID $\tilde{i}$. The model is trained by minimizing the sequence-to-sequence cross-entropy loss with teacher forcing:
\begin{equation}\label{eq:seq2seq_loss}
    \mathcal{L} = - \sum_{t=1}^{|\tilde{i}|} \text{log} P(\tilde{i}^t | C_u, C_v, \tilde{i}^{<t}), \\
\end{equation}
where $\tilde{i}^{t}$ is a $t$-th token of $\tilde{i}$, and $\tilde{i}^{<t}$ denotes the sequence of tokens generated before the $t$-th token.

\subsection{Trend-aware Inference}\label{sec:trend_inference}
We design trend-aware inference to incorporate real-time item trends at recommendation time $t_{|s_u|+1}$ into the final ranking. This training-free method adjusts predictions to reflect dynamic patterns, such as rapidly trending items that emerged after training. This ensures timely recommendations without requiring model retraining, as further demonstrated in Appendix~\ref{sec:app_case_trend}.

\noindent
\textbf{Beam Score}.
The beam score for an item ID $\tilde{i}$ is defined as:
\begin{equation}\label{eq:beam_score}
s_{\text{beam}}(\tilde{i}) = \sum_{t=1}^{|\tilde{i}|} \log P(\tilde{i}^t | C_u, C_v, \tilde{i}^{<t}).
\end{equation}
\noindent
Based on this score, it yields the $B$ most probable item IDs, where $B$ is the beam size. To generate valid IDs, we use a prefix tree Trie~\cite{cormen2022intro2algoTrie}, following existing works~\cite{nips/Tay/DSI, nips/WangHWMWCXCZL0022/NCI}.

\begin{table*}[]\scriptsize
\centering
\setlength{\tabcolsep}{3.32pt}
\renewcommand{\arraystretch}{1.2}
\begin{tabular}{ccccc|cccc|cccc|cccc}
\toprule
\multicolumn{1}{c|}{\multirow{2}{*}{Model}} & \multicolumn{4}{c|}{Beauty} & \multicolumn{4}{c|}{Toys} & \multicolumn{4}{c|}{Sports} & \multicolumn{4}{c}{Yelp} \\
\multicolumn{1}{c|}{} & R@5 & N@5 & R@10 & N@10 & R@5 & N@5 & R@10 & N@10 & R@5 & N@5 & R@10 & N@10 & R@5 & N@5 & R@10 & N@10 \\
\midrule
& \multicolumn{16}{c}{\textit{Traditional recommendation models}} \\
\midrule
\multicolumn{1}{c|}{GRU4Rec} & 0.0429 & 0.0288 & 0.0643 & 0.0357 & 0.0371 & 0.0254 & 0.0549 & 0.0311 & 0.0237 & 0.0154 & 0.0373 & 0.0197 & 0.0240 & 0.0157 & 0.0398 & 0.0207 \\
\multicolumn{1}{c|}{HGN} & 0.0350 & 0.0217 & 0.0589 & 0.0294 & 0.0345 & 0.0212 & 0.0553 & 0.0279 & 0.0203 & 0.0127 & 0.0340 & 0.0171 & 0.0366 & 0.0250 & 0.0532 & 0.0304 \\
\multicolumn{1}{c|}{SASRec} & 0.0323 & 0.0200 & 0.0475 & 0.0249 & 0.0339 & 0.0208 & 0.0442 & 0.0241 & 0.0147 & 0.0089 & 0.0220 & 0.0113 & 0.0284 & 0.0214 & 0.0353 & 0.0245 \\
\multicolumn{1}{c|}{BERT4Rec} & 0.0267 & 0.0165 & 0.0450 & 0.0224 & 0.0210 & 0.0131 & 0.0355 & 0.0178 & 0.0136 & 0.0085 & 0.0233 & 0.0116 & 0.0244 & 0.0159 & 0.0401 & 0.0210 \\
\multicolumn{1}{c|}{FDSA} & 0.0570 & 0.0412 & 0.0777 & 0.0478 & 0.0619 & 0.0455 & 0.0805 & 0.0514 & 0.0283 & 0.0201 & 0.0399 & 0.0238 & 0.0331 & 0.0218 & 0.0534 & 0.0284 \\
\multicolumn{1}{c|}{S$^3$Rec} & 0.0377 & 0.0235 & 0.0627 & 0.0315 & 0.0365 & 0.0231 & 0.0592 & 0.0304 & 0.0229 & 0.0145 & 0.0370 & 0.0190 & 0.0190 & 0.0117 & 0.0321 & 0.0159 \\
\midrule
& \multicolumn{16}{c}{\textit{Temporal recommendation models}} \\
\midrule
\multicolumn{1}{c|}{TiSASRec} & 0.0564 & 0.0359 & 0.0842 & 0.0449 & 0.0665 & 0.0410 & 0.0944 & 0.0499 & 0.0312 & 0.0178 & 0.0474 & 0.0231 & 0.0427 & 0.0323 & 0.0610 & 0.0382 \\
\multicolumn{1}{c|}{TiCoSeRec} & 0.0377 & 0.0186 & 0.0622 & 0.0260 & 0.0408 & 0.0212 & 0.0663 & 0.0292 & 0.0265 & 0.0147 & 0.0455 & 0.0219 & 0.0433 & 0.0301 & 0.0618 & 0.0354 \\
\multicolumn{1}{c|}{HM4SR} & 0.0566 & 0.0409 & 0.0773 & 0.0476 & 0.0647 & 0.0480 & 0.0847 & 0.0545 & 0.0288 & 0.0204 & 0.0402 & 0.0241 & 0.0273 & 0.0185 & 0.0447 & 0.0241 \\
\multicolumn{1}{c|}{HORAE} & 0.0508 & 0.0310 & 0.0834 & 0.0415 & 0.0555 & 0.0331 & 0.0902 & 0.0442 & \ul{0.0379} &  0.0235 & \ul{0.0620} & 0.0313 & 0.0419 & 0.0279 & 0.0663 & 0.0357 \\
\midrule
& \multicolumn{16}{c}{\textit{Generative recommendation models}} \\
\midrule
\multicolumn{1}{c|}{P5-SID} & 0.0465 & 0.0329 & 0.0638 & 0.0384 & 0.0216 & 0.0151 & 0.0325 & 0.0186 & 0.0295 & 0.0212 & 0.0403 & 0.0247 & 0.0299 & 0.0211 & 0.0432 & 0.0253 \\
\multicolumn{1}{c|}{P5-CID} & 0.0465 & 0.0325 & 0.0668 & 0.0391 & 0.0223 & 0.0143 & 0.0357 & 0.0186 & 0.0295 & 0.0214 & 0.0420 & 0.0254 & 0.0226 & 0.0155 & 0.0363 & 0.0199 \\
\multicolumn{1}{c|}{P5-SemID} & 0.0459 & 0.0327 & 0.0667 & 0.0394 & 0.0264 & 0.0178 & 0.0416 & 0.0227 & 0.0336 & 0.0243 & 0.0481 & 0.0290 & 0.0212 & 0.0143 & 0.0329 & 0.0181 \\
\multicolumn{1}{c|}{TIGER} & 0.0352 & 0.0236 & 0.0533 & 0.0294 & 0.0274 & 0.0174 & 0.0438 & 0.0227 & 0.0176 & 0.0111 & 0.0311 & 0.0146 & 0.0164 & 0.0103 & 0.0262 & 0.0135 \\
\multicolumn{1}{c|}{IDGenRec$^\dagger$} & 0.0463 & 0.0328 & 0.0665 & 0.0393 & 0.0462 & 0.0323 & 0.0651 & 0.0383 & 0.0273 & 0.0186 & 0.0403 & 0.0228 & 0.0310 & 0.0219 & 0.0448 & 0.0263 \\
\multicolumn{1}{c|}{ELMRec$^\dagger$} & 0.0372 & 0.0267 & 0.0506 & 0.0310 & 0.0148 & 0.0119 & 0.0193 & 0.0131 & 0.0241 & 0.0181 & 0.0307 & 0.0203 & 0.0424 & 0.0301 & 0.0501 & 0.0324 \\
\multicolumn{1}{c|}{LETTER} & 0.0364 & 0.0243 & 0.0560 & 0.0306 & 0.0309 & 0.0296 & 0.0493 & 0.0262 & 0.0209 & 0.0136 & 0.0331 & 0.0176 & 0.0298 & 0.0218 & 0.0403 & 0.0252 \\
\multicolumn{1}{c|}{LC-Rec} & 0.0503 & 0.0352 & 0.0715 & 0.0420 & 0.0543 & 0.0385 & 0.0753 & 0.0453 & 0.0259 & 0.0175 & 0.0384 & 0.0216 & 0.0341 & 0.0235 & 0.0501 & 0.0286 \\
\multicolumn{1}{c|}{GRAM}  & \ul{0.0641} & \ul{0.0451} & \ul{0.0890} & \ul{0.0531} & \ul{0.0705} & \ul{0.0510} & \ul{0.0958} & \ul{0.0592} & 0.0375 & \ul{0.0256} & 0.0554 & \ul{0.0314} & \ul{0.0476} & \ul{0.0326} & \ul{0.0698} & \ul{0.0397} \\
\midrule
\rowcolor{gray!20} 
\multicolumn{1}{c|}{GRUT} & \textbf{0.0740} & \textbf{0.0511} & \textbf{0.1095} & \textbf{0.0626} & \textbf{0.0769} & \textbf{0.0531} & \textbf{0.1105} & \textbf{0.0640} & \textbf{0.0427} & \textbf{0.0293} & \textbf{0.0626} & \textbf{0.0357} & \textbf{0.0488} & \textbf{0.0346} & \textbf{0.0702} & \textbf{0.0415} \\
\midrule
\multicolumn{1}{c|}{Gain (\%)} & 15.4* & 13.4*& 23.0* & 17.8* & 9.1* & 4.2* & 15.4* & 8.1* & 12.7* & 14.3* & 1.1 & 13.9* & 2.6 & 6.3* & 0.6 & 4.5* \\
\bottomrule
\end{tabular}
\vspace{-1.5mm}
\caption{Overall performance comparison. The best model is marked in \textbf{bold}, and the second-best model is \ul{underlined}. Gain measures the improvement of the proposed method over the best competitive baseline. `$*$' indicates statistical significance $(p < 0.05)$ by a two-tailed $t$-test. `$\dagger$' indicates baselines where results differ from the original papers after addressing preprocessing issues. Please see Appendix~\ref{sec:app_expsetup_detail_mod} for further details.}\label{tab:exp_overall}
\vspace{-5.5mm}
\end{table*}

\noindent
\textbf{Trend Score}.
We calculate the trend score for an item $i$ as follows:
\begin{equation}\label{eq:trend_score}
s_{\text{trend}}(i) = \log(\frac{r_i}{\max_j r_j}+1),
\end{equation}
\noindent
where $r_i$ is the number of appearances of item $i$. The logarithmic scale prevents high values from dominating the score, while normalization by the maximum frequency maintains the relative importance across items. To consider the recent trend, $r_i$ is counted only during the $N$ most recent days before the recommendation time $t_{|s_u|+1}$. The window size $N$ is a hyperparameter that can be adjusted according to the characteristics of the recommendation domain or the trend changes.

\noindent
\textbf{Score Aggregation}.
We aggregate both scores for the final ranking. For $B$ items obtained after beam search, the final score is calculated as:
\begin{equation}\label{eq:final_score}
s_{\text{final}}(i) = s_{\text{beam}}(\tilde{i}) + \lambda \cdot s_{\text{trend}}(i),
\end{equation}
where $\lambda$ is a hyperparameter to control the trend influence. Since trend scores can be pre-computed, it adds minimal computational overhead while adjusting model predictions with trending items. (See Appendix~\ref{sec:app_expresult_latency} for details). Notably, trend-aware inference can be applied to various generative recommenders, as demonstrated in Appendix~\ref{sec:app_expresult_trend}.

\section{Experimental Setup}\label{sec:setup}

\noindent
\textbf{Datasets}. 
We conduct experiments on four real-world datasets: three subcategories from the Amazon review dataset~\cite{sigir/McAuleyTSH15Amazonreview, www/HeM16Amazonreview}\footnote{\url{https://jmcauley.ucsd.edu/data/amazon/}} (``Sports and Outdoors'', ``Beauty'', and ``Toys and Games'') and the Yelp dataset\footnote{\url{https://www.yelp.com/dataset}}. We apply the standard 5-core filtering, removing users and items with fewer than five interactions, following \citet{HuaXGZ23P5Howtoindex}. The data statistics are in Table~\ref{tab:statistics}. 

\noindent
\textbf{Evaluation Protocols and Metrics}. 
We adopt the \emph{leave-one-out} strategy to split train, validation, and test sets following \citet{KangM18SASRec,ZhengHLCZCW24LCRec}. For each user sequence, we use the last item for testing, the second last item as validation data, and the remaining items as training data. Rather than sampling items, we conduct \textit{full-ranking evaluations} on all items to ensure an accurate assessment. For metrics, we adopt top-$k$ Recall (R@$k$) and Normalized Discounted Cumulative Gain (N@$k$) with cutoff $k=\{5, 10\}$.

\noindent
\textbf{Baselines}. 
We validate the effectiveness of~\ours~against the following nineteen sequential recommenders as baselines. 
For traditional baselines, we adopt six models: \textbf{GRU4Rec}~\cite{HidasiKBT15GRU4Rec}, \textbf{HGN}~\cite{kdd/MaKL19HGN}, \textbf{SASRec}~\cite{KangM18SASRec}, \textbf{BERT4Rec}~\cite{SunLWPLOJ19BERT4Rec}, \textbf{FDSA}~\cite{ZhangZLSXWLZ19FDSA}, and \textbf{S$^3$Rec}~\cite{ZhouWZZWZWW20S3Rec}. 
For temporal baselines, we adopt four models: \textbf{TiSASRec}~\cite{LiWM20TISASREC}, \textbf{TiCoSeRec}~\cite{aaai/Dang23TiCoSeRec}, \textbf{HM4SR}~\cite{www/Zhang25HM4SR}, and \textbf{HORAE}~\cite{TOIS/Hu25HORAE}. 
Lastly, we adopt nine state-of-the-art generative recommenders: \textbf{P5-SID}, \textbf{P5-CID}, \textbf{P5-SemID}~\cite{HuaXGZ23P5Howtoindex}, \textbf{TIGER}~\cite{RajputMSKVHHT0S23TIGER}, \textbf{IDGenRec}~\cite{Tan24IDGenRec}, \textbf{ELMRec}~\cite{WangCFS24ELMRec}, \textbf{LETTER}~\cite{cikm24LETTER}, \textbf{LC-Rec}~\cite{ZhengHLCZCW24LCRec}, and \textbf{GRAM}~\cite{acl25GRAM}. The detailed descriptions are in Appendix~\ref{sec:app_expsetup_baseline}.

\noindent
\textbf{Implementation Details}. 
The maximum item sequence length was set to 20, following~\citet{ZhengHLCZCW24LCRec}. We tuned all hyperparameters on the validation set using NDCG@10. We used Adam optimizer~\cite{KingmaB14Adam} with a learning rate of 0.001 and a linear scheduler with a warm-up ratio of 0.05. The maximum text length and the batch size were set to 128. Consistent with the generative baselines~\cite{HuaXGZ23P5Howtoindex, Tan24IDGenRec, WangCFS24ELMRec, acl25GRAM}, we initialized with T5-small~\cite{JMLR/Raffel2020/T5}. Due to space limits, we provide further details in Appendix~\ref{sec:app_expsetup_detail}.

\begin{figure}[t]\small
\centering
\includegraphics[width=1.0\linewidth]{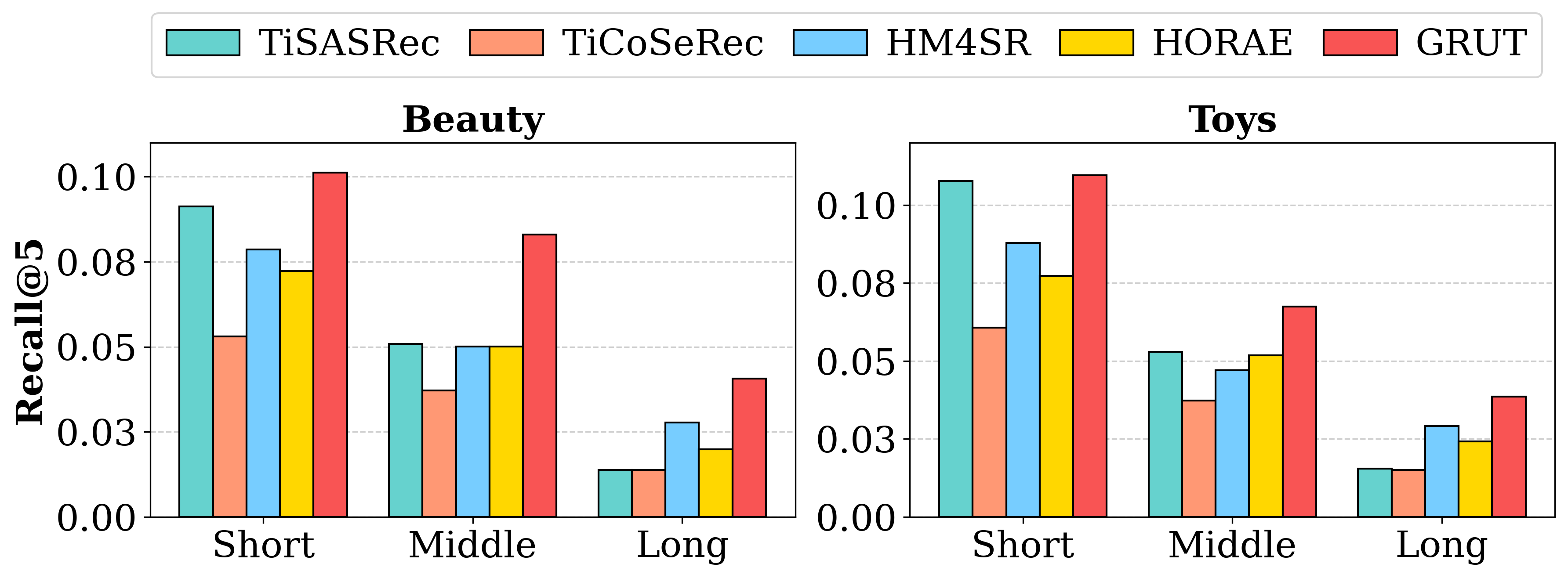}
\includegraphics[width=1.0\linewidth]{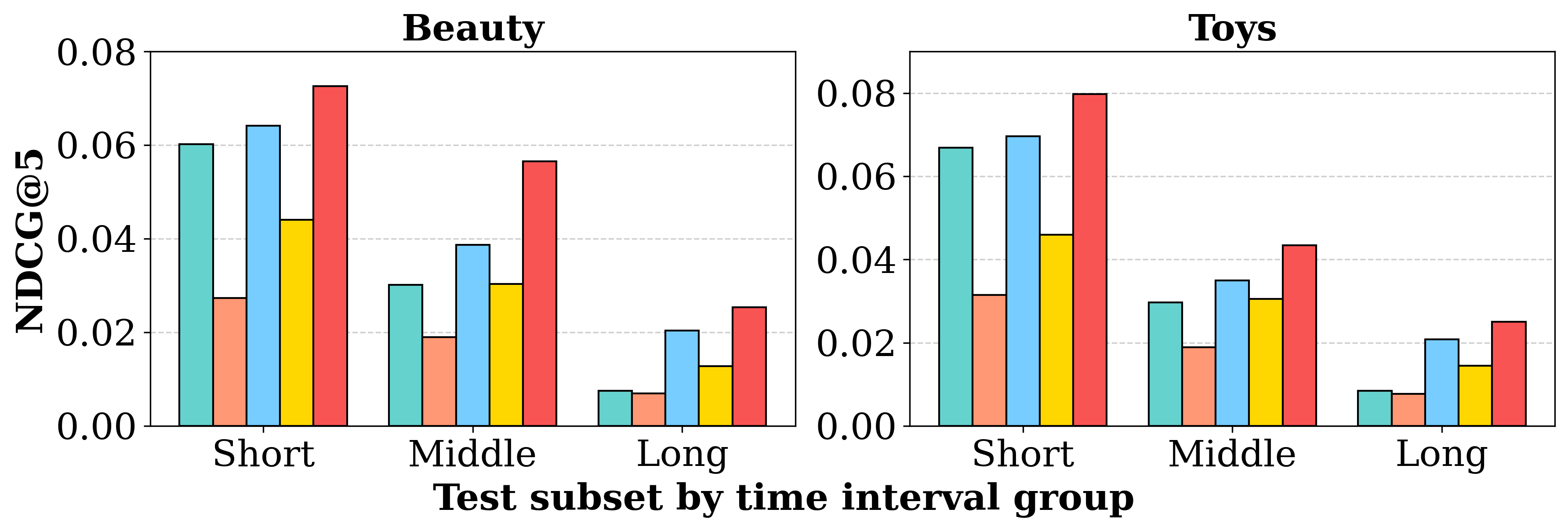}
\vspace{-5.5mm}
\caption{Performance comparison across time interval groups, defined by the number of days between each user’s most recent interaction and the target item.}\label{fig:exp_time_interval_group}
\vspace{-3mm}
\end{figure}

\section{Experimental Results}\label{sec:results}

\subsection{Main Results}\label{sec:exp_main}

\noindent
\textbf{Overall Performance}. 
As shown in Table~\ref{tab:exp_overall}, we thoroughly evaluate the effectiveness of \ours~on four real-world datasets, revealing the following key findings:
(i) \ours~exhibits the state-of-the-art or comparable performance against existing baselines, achieving up to 15.4\% and 14.3\% gains in R@5 and N@5, respectively. \ours~outperforms the best temporal baseline by 30.8\% in R@5 and exceeds the best generative baseline by 15.4\% in R@5. It demonstrates the effectiveness of \ours~in integrating temporal dynamics with generative recommendations.
(ii) Temporal models generally surpass both traditional and generative baselines, highlighting the crucial role of temporal information in capturing evolving user preferences.

\noindent
\textbf{Performance by Time Interval Group}.
Figure~\ref{fig:exp_time_interval_group} illustrates the performance of \ours~and temporal models depending on time intervals between each user's most recent interaction and target item. We categorize users into Short, Middle, and Long subsets.\footnote{Please refer to Appendix~\ref{sec:app_expsetup_data} for detailed statistics.} Our observations are as follows: 
(i) Performance decreases across all models as time intervals increase. It reflects user preference drift over long time intervals between interactions, which presents significant challenges for prediction~\cite{LiWM20TISASREC}\footnote{This is also shown in our analysis in Appendix~\ref{sec:app_expresult_shift}.}.
(ii) \ours~delivers substantial gains in Long interval groups with gains of 32.6--46.0\% in R@5 and 20.2--24.0\% in N@5 compared to the best baseline HM4SR. It confirms the effectiveness of \ours~in identifying preference shifts of users.
(iii) The temporal models that utilize textual metadata (HM4SR, HORAE, \ours) relatively perform better with longer temporal gaps, implying that textual metadata provides valuable signals when recent behavioral items are insufficient.

\begin{table}[]\footnotesize
\vspace{-1.5mm}
\setlength{\tabcolsep}{3.6pt}
\renewcommand{\arraystretch}{0.9}
\begin{tabular}{c|cc|cc}
\toprule
\multirow{2}{*}{Type}    & \multicolumn{2}{c|}{Beauty} & \multicolumn{2}{c}{Toys} \\
            & R@5          & N@5         & R@5         & N@5       \\ \midrule
\textbf{Target-relative + Abs.} & \textbf{0.0740} & \textbf{0.0511} & \textbf{0.0769} & \textbf{0.0531} \\ \midrule
None & 0.0586 & 0.0408 & 0.0583 & 0.0410 \\
Absolute & 0.0605 & 0.0415 & 0.0625 & 0.0428 \\
Relative & 0.0606 & 0.0426 & 0.0588 & 0.0412 \\
Target-relative & 0.0686 & 0.0486 & 0.0692 & 0.0486 \\
Relative + Abs. & 0.0618 & 0.0434 & 0.0631 & 0.0431 \\
\bottomrule
\end{tabular}
\vspace{-2mm}
\caption{Performance of \ours~over time information types in $C_u$. `Abs.' denotes the absolute timestamps.}\label{tab:exp_time_info}

\end{table}
\begin{table}[]\footnotesize
\vspace{-1.5mm}
\setlength{\tabcolsep}{3.6pt}
\renewcommand{\arraystretch}{0.9}
\begin{tabular}{c|cc|cc}
\toprule
\multirow{2}{*}{Model}    & \multicolumn{2}{c|}{Beauty} & \multicolumn{2}{c}{Toys} \\
            & R@5          & N@5         & R@5         & N@5       \\ \midrule
\textbf{G{\scriptsize RUT}} & \textbf{0.0740} & \textbf{0.0511} & \textbf{0.0769} & \textbf{0.0531} \\ \midrule
w/o user-level & 0.0586 & 0.0408 & 0.0583 & 0.0410 \\
w/o item-level & 0.0713 & 0.0492 & 0.0755 & 0.0518 \\
w/o trend score ($\lambda$ = 0) & 0.0726 & 0.0506 & 0.0758 & 0.0526 \\
\midrule
w/o context embedding & 0.0711 & 0.0500 & 0.0728 & 0.0524 \\
w/o epsilon ($\epsilon$ = 1) & 0.0681 & 0.0486 & 0.0732 & 0.0529 \\
\bottomrule
\end{tabular}
\vspace{-1mm}
\caption{Ablation study of \ours. We examine the effect of (i) time-aware prompting, (ii) trend-aware inference, and (iii) additional techniques. }\label{tab:exp_ablation}
\vspace{-3mm}
\end{table}

\newcolumntype{C}[1]{>{\centering\arraybackslash}m{#1\columnwidth}}

\begin{table}[t]\scriptsize
\centering
\setlength{\tabcolsep}{2pt}
\renewcommand{\arraystretch}{0.3}

\begin{tabular}{C{0.125}|*{5}{C{0.15}}} 
\toprule
\multicolumn{6}{c}{\textbf{User sequence (\valfont{ASIN:A1M2CZP3XOVZO5})}} \\ \midrule

\textbf{Image} &
\cellimg{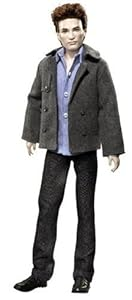} &
\cellimg{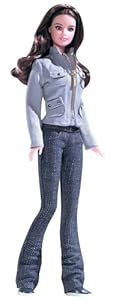} &
\cellimg{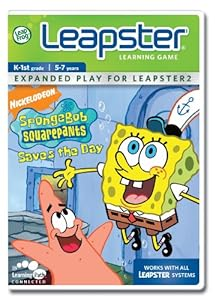} &
\cellimg{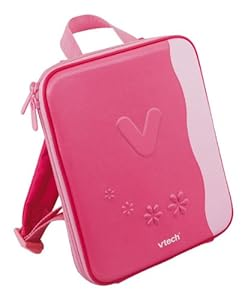} &
\cellimg{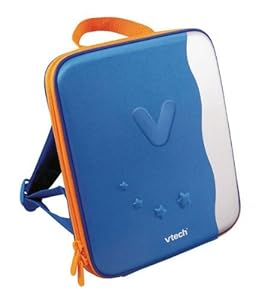} \\ \midrule

\textbf{Name}     & \valfont{\shortstack{Edward\\Doll}}
                 & \valfont{\shortstack{Bella\\Doll}}
                 & \valfont{\shortstack{SpongeBob\\Game}}
                 & \valfont{\shortstack{InnoTab\\Storage (Pink)}}
                 & \valfont{\shortstack{InnoTab\\Storage (Blue)}} \\ \midrule
\textbf{Category} & \valfont{Dolls}
                 & \valfont{Dolls}
                 & \valfont{\shortstack{Learning\\Game}}
                 & \valfont{\shortstack{System\\Acc.}}
                 & \valfont{\shortstack{System\\Acc.}} \\ \midrule
\textbf{Time}     & \valfont{2010-01-11} & \valfont{2010-01-11} & \valfont{2010-02-16} & \valfont{2012-12-11} & \valfont{2012-12-11} \\ \bottomrule


\multicolumn{6}{c}{} \\[0.4em] 
\rowcolor{gray!20} 
\multicolumn{6}{c}{\textbf{GRUT Top-5 predictions at 2012-12-11 (Without temporal information)}} \\ \midrule

\textbf{Ranking} & 1 & 2 & 3 & 4 & 5 \\ \midrule

\textbf{Image} &
\cellimg{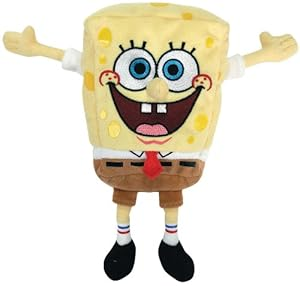} &
\cellimg{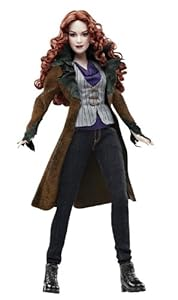} &
\cellimg{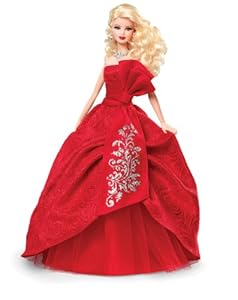} &
\cellimg{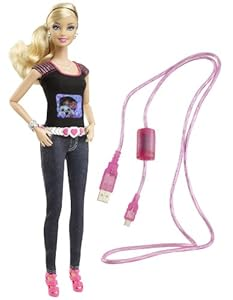} &
\cellimg{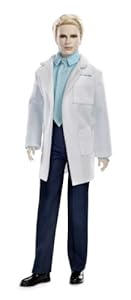} \\ \midrule

\textbf{Name}     & \valfont{\shortstack{SpongeBob\\Beanie}}
                 & \valfont{\shortstack{Eclipse\\Victoria Doll}}
                 & \valfont{\shortstack{2012 Holiday\\Doll}}
                 & \valfont{\shortstack{Photo Fashion\\Doll}}
                 & \valfont{\shortstack{Carlisle\\Doll}} \\ \midrule
\textbf{Category} & \valfont{Plush}
                 & \valfont{Dolls}
                 & \valfont{Dolls}
                 & \valfont{Dolls}
                 & \valfont{Dolls} \\ \bottomrule

\multicolumn{6}{c}{} \\[0.4em] 
\rowcolor{gray!20} 
\multicolumn{6}{c}{\textbf{GRUT Top-5 predictions at 2012-12-11 (With temporal information)}} \\ \midrule

\textbf{Ranking} & 1 & 2 & 3 & 4 & 5 \\ \midrule

\textbf{Image} &
\cellimg{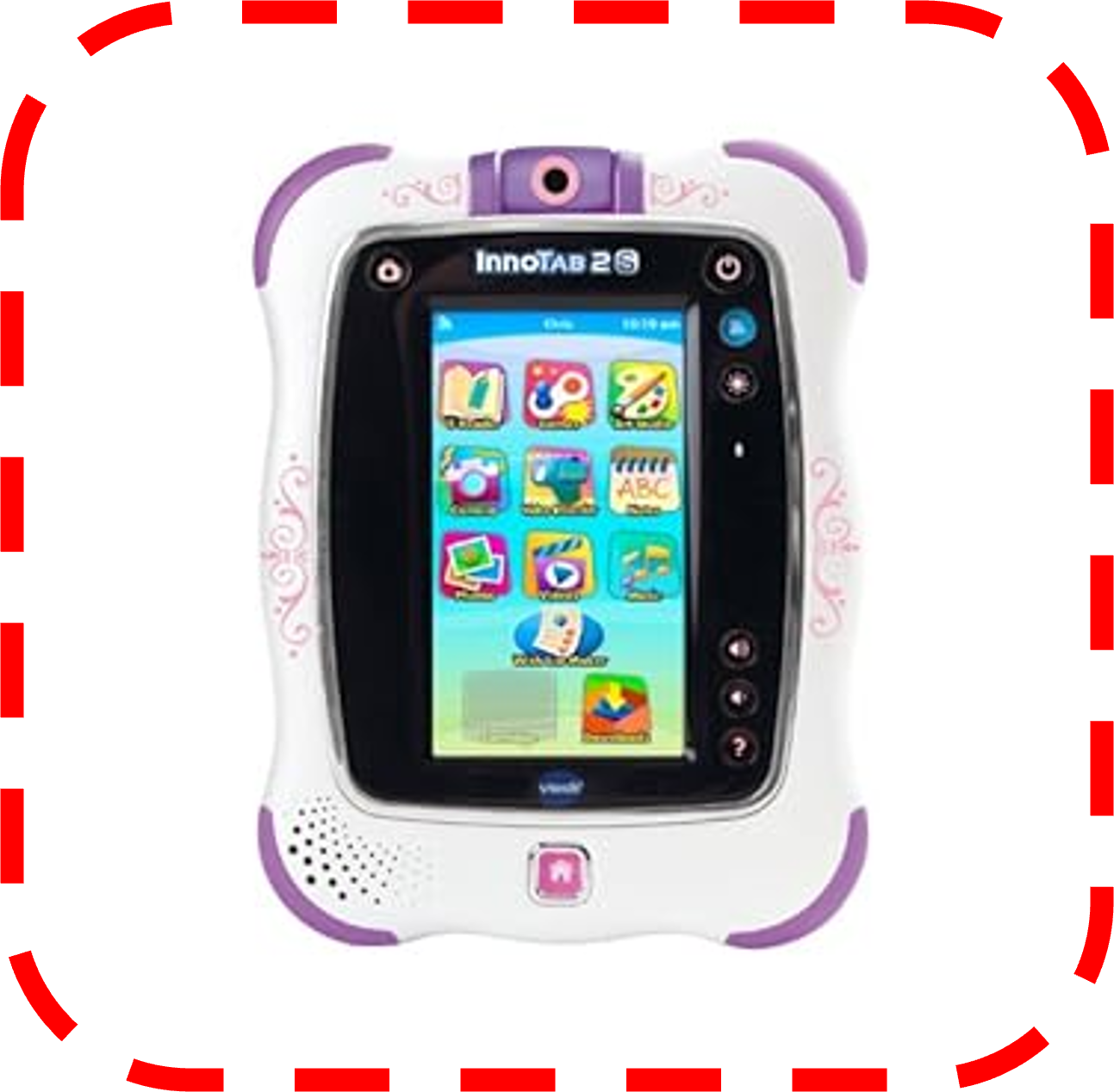} &
\cellimg{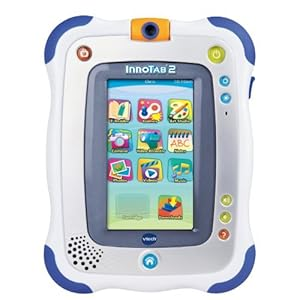} &
\cellimg{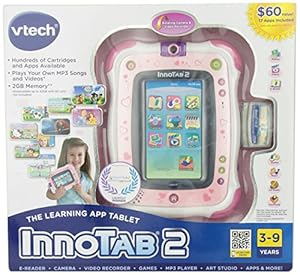} &
\cellimg{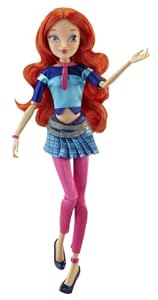} &
\cellimg{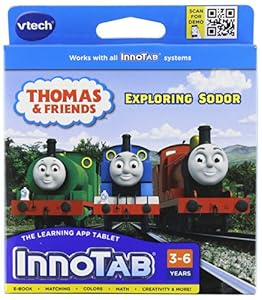} \\ \midrule

\textbf{Name}     & \valfont{\shortstack{InnoTab 2S\\Tablet}}
                 & \valfont{\shortstack{InnoTab 2\\White Tablet}}
                 & \valfont{\shortstack{InnoTab 2\\Pink Tablet}}
                 & \valfont{\shortstack{Winx Bloom\\Doll}}
                 & \valfont{\shortstack{InnoTab\\Thomas}} \\ \midrule
\textbf{Category} & \valfont{\shortstack{Learning\\Tablet}}
                 & \valfont{\shortstack{Learning\\Tablet}}
                 & \valfont{\shortstack{Learning\\Tablet}}
                 & \valfont{Dolls}
                 & \valfont{\shortstack{Learning\\Software}} \\ \bottomrule

\end{tabular}

\caption{\ours’s top-5 predictions on the Toys dataset with and without temporal information. The five most recent items in the sequence are shown for simplicity. The target item is marked with a red dotted line.}\label{fig:case_study_using_time_case2}
\vspace{-3.5mm}
\end{table}

\subsection{Ablation Study}

\noindent
\textbf{Effect of Time Information Types}. 
Table~\ref{tab:exp_time_info} presents the impact of temporal information types in the user-level temporal context $C_u$. We compare six variants: None, Absolute timestamps ($t_i$), Relative intervals ($t_{i+1}-t_i$), Target-relative intervals ($t_{|s_u|+1}-t_i$), Relative + Absolute, and Target-relative + Absolute\footnote{See Appendix~\ref{sec:app_expsetup_ablation_input} for detailed prompts of each variant.}. All time-aware variants outperform the baseline, with up to 31.9\% gains in R@5, confirming the benefits of verbalizing temporal dynamics. The target-relative intervals especially achieve the highest performance, suggesting that recency relative to recommendation time effectively captures user preferences. Notably, combining absolute timestamps and interval information consistently yields gains of 2.0\%--11.1\% in R@5. It demonstrates that two distinct forms of temporal signals successfully complement each other.

\noindent
\textbf{Effect of Various Components}. 
Table~\ref{tab:exp_ablation} shows the effectiveness of various components in \ours. 
(i) Both user-level temporal context $C_u$ and item-level transition context $C_v$ contribute to performance. Specifically, temporal information in $C_u$ enhances R@5 by up to 31.9\%. It highlights the importance of user-specific temporal patterns, while transition patterns also convey valuable additional guidance. 
(ii) Trend-aware inference not only provides flexibility in controlling trend influence but also improves recommendation accuracy by up to 1.9\% in R@5. This improvement results from incorporating real-time trend signals that were unavailable during training.
(iii) The context-type embeddings $\mathbf{P}$ in Eq.~\eqref{eq:context_emb} and $\epsilon$ in Eq.~\eqref{eq:key_value_matrix} boost R@5 by up to 5.6\% and 8.6\%, respectively. It indicates that distinguishing context types while ensuring transitions as supplementary information enhances recommendation accuracy.

\subsection{In-depth Analysis}\label{sec:exp_in_depth}

\begin{figure}[t]\small
\centering
\includegraphics[width=1.0\linewidth]{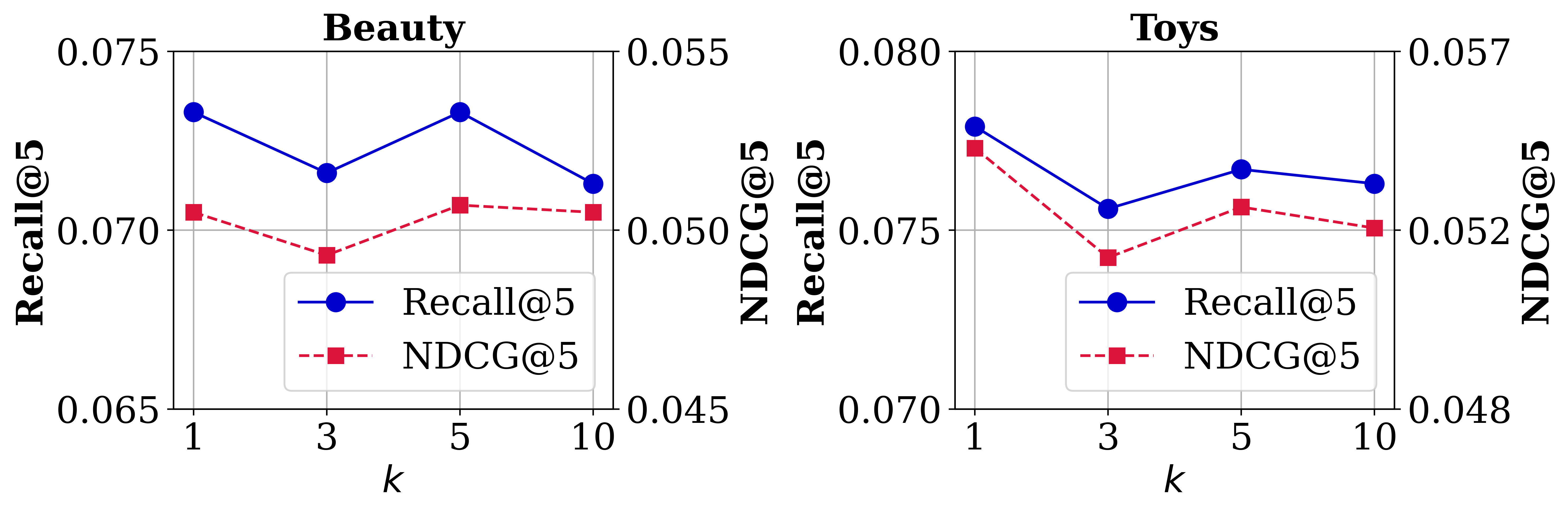}
\vspace{-4.5mm}
\caption{Performance of \ours~over varying the number of neighboring items $k$ in $C_v$.}\label{fig:exp_hyper_neighbor}
\vspace{-3.5mm}
\end{figure}

\begin{figure}[t]\small
\centering
\includegraphics[width=1.0\linewidth]{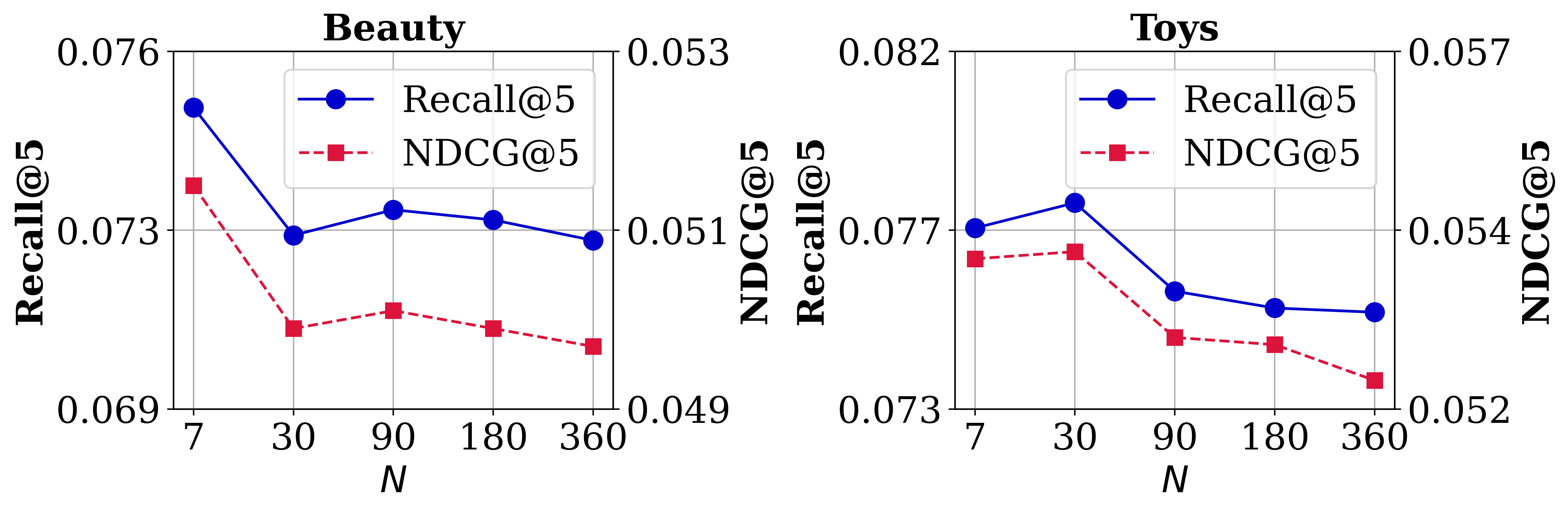}
\vspace{-4.5mm}
\caption{Performance of \ours~over varying the window size $N$ in the trend score.}\label{fig:exp_hyper_window}
\vspace{-6.5mm}
\end{figure}

\noindent
\textbf{Case Study}.
Table~\ref{fig:case_study_using_time_case2} illustrates the impact of temporal information on the recommendation results of \ours. Without temporal information, the model recommends `\textit{Plush}' and `\textit{Dolls}', missing that the user's purchasing pattern has shifted over the past two years from `\textit{Dolls}' to `\textit{InnoTab}'. Conversely, \ours~with temporal information successfully identifies the preference shift and recommends an `\textit{InnoTab 2S Tablet}', while also suggesting a `\textit{Winx Bloom Doll}'. It depicts that temporal dynamics are crucial in capturing user preferences that evolve over time, leading to more accurate recommendations. Please see Appendix~\ref{sec:app_case} for additional cases.

\noindent
\textbf{Hyperparameter Sensitivity}.
Figures~\ref{fig:exp_hyper_neighbor} and \ref{fig:exp_hyper_window} show the performance of \ours~when varying neighboring items $k$ and trend window size $N$. We observe optimal performance at $k$ = 1 for both Beauty and Toys datasets, suggesting that more neighbors may introduce noise. For $N$, the optimal values for Beauty and Toys are 7 and 30, respectively. It highlights the importance of adjusting the trend window size according to how rapidly preferences change in each domain. An additional analysis of $\epsilon$ and $L$ are in Appendix~\ref{sec:app_expresult_hyper}.

\section{Conclusion}\label{sec:conclusion}

We propose \ours, a novel model that effectively incorporates temporal dynamics into GR. Our time-aware prompting captures both user-specific temporal patterns and item-level transition knowledge. Additionally, trend-aware inference enhances rankings by injecting trend information. Extensive experiments on four benchmark datasets demonstrate improvements of \ours~compared to state-of-the-art recommendation models, up to 15.4\% in R@5 and 14.3\% in N@5, particularly in scenarios with long time intervals between interactions. Our work highlights the importance of time awareness in GR, opening new directions for future models that better reflect evolving user preferences.

\section{Limitations}\label{sec:limitation}
The limitations of our work are as follows. (i) To construct the item-level transition context $C_v$, we include all transition pairs from the training data in a global item transition graph. This approach has a limitation as it may incorporate noise or spurious patterns, \eg, accidental clicks. This challenge has also been noted in previous work~\cite{aaai/Zhang24TGCL4SR}, and future research could apply denoising techniques to extract only meaningful temporal patterns. (ii) Our method currently incorporates temporal information uniformly across all users. However, as pointed out in the existing work~\cite{HLGQZHT23}, users exhibit diverse purchasing patterns which our approach does not explicitly model, presenting another limitation of our work. We believe that modeling user preferences in a user-adaptive manner would be meaningful. For instance, in trend-aware inference, the value of $\lambda$ could be dynamically adjusted according to individual patterns. We leave further exploration as future work.

\section*{Ethics Statement}

This work fully complies with the ACL Ethics Policy. We declare that there are no ethical issues in this paper. The scientific artifacts we have utilized are publicly available for research under permissive licenses, and the utilization of these tools is consistent with their intended applications.

\section*{Acknowledgments} 
This work was supported by the Institute of Information \& communications Technology Planning \& Evaluation (IITP) grant and the National Research Foundation of Korea (NRF) grant funded by the Korea government (MSIT) (No. NRF-RS-2025-00564083, IITP-RS-2019-II190421, IITP-RS-2022-II220680, IITP-2025-RS-2020-II201821).

\bibliography{references}

\begin{thebibliography}{59}
\providecommand{\natexlab}[1]{#1}

\bibitem[{Bao et~al.(2023)Bao, Zhang, Zhang, Wang, Feng, and He}]{BaoZZWF023TALLRec}
Keqin Bao, Jizhi Zhang, Yang Zhang, Wenjie Wang, Fuli Feng, and Xiangnan He. 2023.
\newblock \href {https://doi.org/10.1145/3604915.3608857} {Tallrec: An effective and efficient tuning framework to align large language model with recommendation}.
\newblock In \emph{RecSys}, pages 1007--1014.

\bibitem[{Chen et~al.(2022)Chen, Liu, Li, McAuley, and Xiong}]{ChenLLMX22ICLRec}
Yongjun Chen, Zhiwei Liu, Jia Li, Julian~J. McAuley, and Caiming Xiong. 2022.
\newblock \href {https://doi.org/10.1145/3485447.3512090} {Intent contrastive learning for sequential recommendation}.
\newblock In \emph{WWW}, pages 2172--2182.

\bibitem[{Cho et~al.(2020)Cho, Park, and Yoo}]{recsys/Cho20MEANTIME}
Sung~Min Cho, Eunhyeok Park, and Sungjoo Yoo. 2020.
\newblock \href {https://doi.org/10.1145/3383313.3412216} {{MEANTIME:} mixture of attention mechanisms with multi-temporal embeddings for sequential recommendation}.
\newblock In \emph{RecSys}, pages 515--520.

\bibitem[{Chu et~al.(2024)Chu, Wang, Zhang, Ji, Wang, and Sun}]{corr/24Tempura}
Zhendong Chu, Zichao Wang, Ruiyi Zhang, Yangfeng Ji, Hongning Wang, and Tong Sun. 2024.
\newblock \href {https://doi.org/10.48550/arXiv.2405.02778} {Improve temporal awareness of llms for sequential recommendation}.
\newblock \emph{CoRR}, abs/2405.02778.

\bibitem[{Cormen et~al.(2022)Cormen, Leiserson, Rivest, and Stein}]{cormen2022intro2algoTrie}
Thomas~H Cormen, Charles~E Leiserson, Ronald~L Rivest, and Clifford Stein. 2022.
\newblock \href {https://mitpress.mit.edu/9780262046305/introduction-to-algorithms/} {\emph{Introduction to algorithms}}.
\newblock MIT press.

\bibitem[{Dang et~al.(2023)Dang, Yang, Guo, Jiang, Wang, Xu, Sun, and Liu}]{aaai/Dang23TiCoSeRec}
Yizhou Dang, Enneng Yang, Guibing Guo, Linying Jiang, Xingwei Wang, Xiaoxiao Xu, Qinghui Sun, and Hong Liu. 2023.
\newblock \href {https://doi.org/10.1609/aaai.v37i4.25540} {Uniform sequence better: Time interval aware data augmentation for sequential recommendation}.
\newblock In \emph{AAAI}, pages 4225--4232.

\bibitem[{Fan et~al.(2021)Fan, Liu, Zhang, Xiong, Zheng, and Yu}]{cikm/Fan21TGSRec}
Ziwei Fan, Zhiwei Liu, Jiawei Zhang, Yun Xiong, Lei Zheng, and Philip~S. Yu. 2021.
\newblock \href {https://doi.org/10.1145/3459637.3482242} {Continuous-time sequential recommendation with temporal graph collaborative transformer}.
\newblock In \emph{CIKM}, pages 433--442.

\bibitem[{Geng et~al.(2022)Geng, Liu, Fu, Ge, and Zhang}]{Geng0FGZ22P5}
Shijie Geng, Shuchang Liu, Zuohui Fu, Yingqiang Ge, and Yongfeng Zhang. 2022.
\newblock \href {https://doi.org/10.1145/3523227.3546767} {Recommendation as language processing {(RLP):} {A} unified pretrain, personalized prompt {\&} predict paradigm {(P5)}}.
\newblock In \emph{RecSys}, pages 299--315.

\bibitem[{He and McAuley(2016)}]{www/HeM16Amazonreview}
Ruining He and Julian~J. McAuley. 2016.
\newblock \href {https://doi.org/10.1145/2872427.2883037} {Ups and downs: Modeling the visual evolution of fashion trends with one-class collaborative filtering}.
\newblock In \emph{WWW}, pages 507--517.

\bibitem[{He et~al.(2023)He, Liu, Guo, Qin, Zhang, Hu, and Tang}]{HLGQZHT23}
Zhicheng He, Weiwen Liu, Wei Guo, Jiarui Qin, Yingxue Zhang, Yaochen Hu, and Ruiming Tang. 2023.
\newblock \href {https://doi.org/10.24963/ijcai.2023/746} {A survey on user behavior modeling in recommender systems}.
\newblock In \emph{IJCAI}, pages 6656--6664.

\bibitem[{Hidasi et~al.(2016)Hidasi, Karatzoglou, Baltrunas, and Tikk}]{HidasiKBT15GRU4Rec}
Bal{\'{a}}zs Hidasi, Alexandros Karatzoglou, Linas Baltrunas, and Domonkos Tikk. 2016.
\newblock \href {http://arxiv.org/abs/1511.06939} {Session-based recommendations with recurrent neural networks}.
\newblock In \emph{ICLR}.

\bibitem[{Hu et~al.(2025)Hu, Wu, Tang, Huan, Wang, Zhang, Zhou, Zou, and Li}]{TOIS/Hu25HORAE}
Shirui Hu, Weichang Wu, Zuoli Tang, Zhaoxin Huan, Lin Wang, Xiaolu Zhang, Jun Zhou, Lixin Zou, and Chenliang Li. 2025.
\newblock \href {https://doi.org/10.1145/3727645} {Horae: Temporal multi-interest pre-training for sequential recommendation}.
\newblock \emph{ACM Trans. Inf. Syst.}

\bibitem[{Hua et~al.(2023)Hua, Xu, Ge, and Zhang}]{HuaXGZ23P5Howtoindex}
Wenyue Hua, Shuyuan Xu, Yingqiang Ge, and Yongfeng Zhang. 2023.
\newblock \href {https://doi.org/10.1145/3624918.3625339} {How to index item ids for recommendation foundation models}.
\newblock In \emph{SIGIR-AP}, pages 195--204.

\bibitem[{Izacard and Grave(2021)}]{eacl/IzacardG21/FiD}
Gautier Izacard and Edouard Grave. 2021.
\newblock \href {https://doi.org/10.18653/v1/2021.eacl-main.74} {Leveraging passage retrieval with generative models for open domain question answering}.
\newblock In \emph{EACL}, pages 874--880.

\bibitem[{Jones(2004)}]{Jones2004/TFIDF}
Karen~Sparck Jones. 2004.
\newblock \href {https://doi.org/10.1108/00220410410560573} {A statistical interpretation of term specificity and its application in retrieval}.
\newblock \emph{J. Documentation}, 60(5):493--502.

\bibitem[{Kang and McAuley(2018)}]{KangM18SASRec}
Wang{-}Cheng Kang and Julian~J. McAuley. 2018.
\newblock \href {https://doi.org/10.1109/ICDM.2018.00035} {Self-attentive sequential recommendation}.
\newblock In \emph{ICDM}, pages 197--206.

\bibitem[{Kim et~al.(2024)Kim, Kang, Choi, Kim, Yang, and Park}]{KimKC0YP24A-LLMRec}
Sein Kim, Hongseok Kang, Seungyoon Choi, Donghyun Kim, Min{-}Chul Yang, and Chanyoung Park. 2024.
\newblock \href {https://doi.org/10.1145/3637528.3671931} {Large language models meet collaborative filtering: An efficient all-round llm-based recommender system}.
\newblock In \emph{KDD}, pages 1395--1406.

\bibitem[{Kingma and Ba(2015)}]{KingmaB14Adam}
Diederik~P. Kingma and Jimmy Ba. 2015.
\newblock \href {http://arxiv.org/abs/1412.6980} {Adam: {A} method for stochastic optimization}.
\newblock In \emph{ICLR}.

\bibitem[{Lee et~al.(2024)Lee, Roy, Xu, Raiman, Shoeybi, Catanzaro, and Ping}]{abs-2405-17428/NVEmbed}
Chankyu Lee, Rajarshi Roy, Mengyao Xu, Jonathan Raiman, Mohammad Shoeybi, Bryan Catanzaro, and Wei Ping. 2024.
\newblock \href {https://doi.org/10.48550/arXiv.2405.17428} {Nv-embed: Improved techniques for training llms as generalist embedding models}.
\newblock \emph{CoRR}, abs/2405.17428.

\bibitem[{Lee et~al.(2025)Lee, Choi, Choi, Kim, and Lee}]{acl25GRAM}
Sunkyung Lee, Minjin Choi, Eunseong Choi, Hye{-}young Kim, and Jongwuk Lee. 2025.
\newblock \href {https://aclanthology.org/2025.acl-long.1596/} {{GRAM:} generative recommendation via semantic-aware multi-granular late fusion}.
\newblock In \emph{ACL}, pages 33294--33312.

\bibitem[{Lee et~al.(2023)Lee, Choi, and Lee}]{emnlp23GLEN}
Sunkyung Lee, Minjin Choi, and Jongwuk Lee. 2023.
\newblock \href {https://doi.org/10.18653/v1/2023.emnlp-main.477} {{GLEN:} generative retrieval via lexical index learning}.
\newblock In \emph{EMNLP}, pages 7693--7704.

\bibitem[{Li et~al.(2023{\natexlab{a}})Li, Wang, Li, Fu, Shen, Shang, and McAuley}]{LiWLFSSM23Recformer}
Jiacheng Li, Ming Wang, Jin Li, Jinmiao Fu, Xin Shen, Jingbo Shang, and Julian~J. McAuley. 2023{\natexlab{a}}.
\newblock \href {https://doi.org/10.1145/3580305.3599519} {Text is all you need: Learning language representations for sequential recommendation}.
\newblock In \emph{KDD}, pages 1258--1267.

\bibitem[{Li et~al.(2020)Li, Wang, and McAuley}]{LiWM20TISASREC}
Jiacheng Li, Yujie Wang, and Julian~J. McAuley. 2020.
\newblock \href {https://doi.org/10.1145/3336191.3371786} {Time interval aware self-attention for sequential recommendation}.
\newblock In \emph{WSDM}, pages 322--330.

\bibitem[{Li et~al.(2023{\natexlab{b}})Li, Chen, Zhao, Zhang, and Xing}]{LCZZX23E4SRec}
Xinhang Li, Chong Chen, Xiangyu Zhao, Yong Zhang, and Chunxiao Xing. 2023{\natexlab{b}}.
\newblock \href {https://doi.org/10.48550/arXiv.2312.02443} {E4srec: An elegant effective efficient extensible solution of large language models for sequential recommendation}.
\newblock \emph{CoRR}, abs/2312.02443.

\bibitem[{Li et~al.(2024)Li, Lin, Wang, Feng, Pang, Li, Nie, He, and Chua}]{SurveyGenSearchRecSys}
Yongqi Li, Xinyu Lin, Wenjie Wang, Fuli Feng, Liang Pang, Wenjie Li, Liqiang Nie, Xiangnan He, and Tat{-}Seng Chua. 2024.
\newblock \href {https://doi.org/10.48550/arXiv.2404.16924} {A survey of generative search and recommendation in the era of large language models}.
\newblock \emph{CoRR}.

\bibitem[{Liao et~al.(2024)Liao, Li, Yang, Wu, Yuan, Wang, and He}]{sigir/Liao24LLaRA}
Jiayi Liao, Sihang Li, Zhengyi Yang, Jiancan Wu, Yancheng Yuan, Xiang Wang, and Xiangnan He. 2024.
\newblock \href {https://doi.org/10.1145/3626772.3657690} {Llara: Large language-recommendation assistant}.
\newblock In \emph{SIGIR}, pages 1785--1795.

\bibitem[{Lin et~al.(2024)Lin, Wang, Li, Feng, Ng, and Chua}]{kdd24TransRec}
Xinyu Lin, Wenjie Wang, Yongqi Li, Fuli Feng, See{-}Kiong Ng, and Tat{-}Seng Chua. 2024.
\newblock \href {https://doi.org/10.1145/3637528.3671884} {Bridging items and language: {A} transition paradigm for large language model-based recommendation}.
\newblock In \emph{KDD}, pages 1816--1826.

\bibitem[{Liu et~al.(2025)Liu, Wu, Wang, Wang, Zhu, Zhao, Tian, and Zheng}]{WZ00025LLMEmb}
Qidong Liu, Xian Wu, Wanyu Wang, Yejing Wang, Yuanshao Zhu, Xiangyu Zhao, Feng Tian, and Yefeng Zheng. 2025.
\newblock \href {https://doi.org/10.1609/aaai.v39i11.33327} {Llmemb: Large language model can be a good embedding generator for sequential recommendation}.
\newblock In \emph{AAAI}, pages 12183--12191.

\bibitem[{Liu et~al.(2024)Liu, Wu, Wang, Zhang, Tian, Zheng, and Zhao}]{Liu0WLLM-ESR}
Qidong Liu, Xian Wu, Yejing Wang, Zijian Zhang, Feng Tian, Yefeng Zheng, and Xiangyu Zhao. 2024.
\newblock \href {http://papers.nips.cc/paper\_files/paper/2024/hash/2f0728449cb3150189d765fc87afc913-Abstract-Conference.html} {{LLM-ESR:} large language models enhancement for long-tailed sequential recommendation}.
\newblock In \emph{NeurIPS}.

\bibitem[{Ma et~al.(2019)Ma, Kang, and Liu}]{kdd/MaKL19HGN}
Chen Ma, Peng Kang, and Xue Liu. 2019.
\newblock \href {https://doi.org/10.1145/3292500.3330984} {Hierarchical gating networks for sequential recommendation}.
\newblock In \emph{KDD}, pages 825--833.

\bibitem[{McAuley et~al.(2015)McAuley, Targett, Shi, and van~den Hengel}]{sigir/McAuleyTSH15Amazonreview}
Julian~J. McAuley, Christopher Targett, Qinfeng Shi, and Anton van~den Hengel. 2015.
\newblock \href {https://doi.org/10.1145/2766462.2767755} {Image-based recommendations on styles and substitutes}.
\newblock In \emph{SIGIR}, pages 43--52.

\bibitem[{Ni et~al.(2022)Ni, {\'{A}}brego, Constant, Ma, Hall, Cer, and Yang}]{acl/NiACMHCY22/SentenceT5}
Jianmo Ni, Gustavo~Hern{\'{a}}ndez {\'{A}}brego, Noah Constant, Ji~Ma, Keith~B. Hall, Daniel Cer, and Yinfei Yang. 2022.
\newblock \href {https://doi.org/10.18653/v1/2022.findings-acl.146} {Sentence-t5: Scalable sentence encoders from pre-trained text-to-text models}.
\newblock In \emph{Findings of the ACL}, pages 1864--1874.

\bibitem[{Park et~al.(2025)Park, Yoon, Choi, and Lee}]{wsdm/Park25TALE}
Seongmin Park, Mincheol Yoon, Minjin Choi, and Jongwuk Lee. 2025.
\newblock \href {https://doi.org/10.1145/3701551.3703554} {Temporal linear item-item model for sequential recommendation}.
\newblock In \emph{WSDM}, pages 354--362.

\bibitem[{Raffel et~al.(2020)Raffel, Shazeer, Roberts, Lee, Narang, Matena, Zhou, Li, and Liu}]{JMLR/Raffel2020/T5}
Colin Raffel, Noam Shazeer, Adam Roberts, Katherine Lee, Sharan Narang, Michael Matena, Yanqi Zhou, Wei Li, and Peter~J. Liu. 2020.
\newblock \href {https://jmlr.org/papers/v21/20-074.html} {Exploring the limits of transfer learning with a unified text-to-text transformer}.
\newblock \emph{J. Mach. Learn. Res.}, 21:140:1--140:67.

\bibitem[{Rajput et~al.(2023)Rajput, Mehta, Singh, Keshavan, Vu, Heldt, Hong, Tay, Tran, Samost, Kula, Chi, and Sathiamoorthy}]{RajputMSKVHHT0S23TIGER}
Shashank Rajput, Nikhil Mehta, Anima Singh, Raghunandan~Hulikal Keshavan, Trung Vu, Lukasz Heldt, Lichan Hong, Yi~Tay, Vinh~Q. Tran, Jonah Samost, Maciej Kula, Ed~H. Chi, and Mahesh Sathiamoorthy. 2023.
\newblock \href {http://papers.nips.cc/paper\_files/paper/2023/hash/20dcab0f14046a5c6b02b61da9f13229-Abstract-Conference.html} {Recommender systems with generative retrieval}.
\newblock In \emph{NeurIPS}, pages 10299--10315.

\bibitem[{Ren et~al.(2024)Ren, Wei, Xia, Su, Cheng, Wang, Yin, and Huang}]{RenWXSCWY024RLMRec}
Xubin Ren, Wei Wei, Lianghao Xia, Lixin Su, Suqi Cheng, Junfeng Wang, Dawei Yin, and Chao Huang. 2024.
\newblock \href {https://doi.org/10.1145/3589334.3645458} {Representation learning with large language models for recommendation}.
\newblock In \emph{WWW}.

\bibitem[{Sennrich et~al.(2016)Sennrich, Haddow, and Birch}]{Sennrich16aSentencePiece}
Rico Sennrich, Barry Haddow, and Alexandra Birch. 2016.
\newblock \href {https://doi.org/10.18653/v1/p16-1162} {Neural machine translation of rare words with subword units}.
\newblock In \emph{ACL}.

\bibitem[{Sheng et~al.(2025)Sheng, Zhang, Zhang, Chen, Wang, and Chua}]{SZZCWC25AlphaRec}
Leheng Sheng, An~Zhang, Yi~Zhang, Yuxin Chen, Xiang Wang, and Tat{-}Seng Chua. 2025.
\newblock \href {https://openreview.net/forum?id=eIJfOIMN9z} {Language representations can be what recommenders need: Findings and potentials}.
\newblock In \emph{ICLR}.

\bibitem[{Sun et~al.(2019)Sun, Liu, Wu, Pei, Lin, Ou, and Jiang}]{SunLWPLOJ19BERT4Rec}
Fei Sun, Jun Liu, Jian Wu, Changhua Pei, Xiao Lin, Wenwu Ou, and Peng Jiang. 2019.
\newblock \href {https://doi.org/10.1145/3357384.3357895} {Bert4rec: Sequential recommendation with bidirectional encoder representations from transformer}.
\newblock In \emph{CIKM}, pages 1441--1450.

\bibitem[{Tan et~al.(2024)Tan, Xu, Hua, Ge, Li, and Zhang}]{Tan24IDGenRec}
Juntao Tan, Shuyuan Xu, Wenyue Hua, Yingqiang Ge, Zelong Li, and Yongfeng Zhang. 2024.
\newblock \href {https://doi.org/10.1145/3626772.3657821} {Idgenrec: Llm-recsys alignment with textual id learning}.
\newblock In \emph{SIGIR}, page 355–364.

\bibitem[{Tang and Wang(2018)}]{TangW18caser}
Jiaxi Tang and Ke~Wang. 2018.
\newblock \href {https://doi.org/10.1145/3159652.3159656} {Personalized top-n sequential recommendation via convolutional sequence embedding}.
\newblock In \emph{WSDM}, pages 565--573.

\bibitem[{Tay et~al.(2022)Tay, Tran, Dehghani, Ni, Bahri, Mehta, Qin, Hui, Zhao, Gupta, Schuster, Cohen, and Metzler}]{nips/Tay/DSI}
Yi~Tay, Vinh Tran, Mostafa Dehghani, Jianmo Ni, Dara Bahri, Harsh Mehta, Zhen Qin, Kai Hui, Zhe Zhao, Jai~Prakash Gupta, Tal Schuster, William~W. Cohen, and Donald Metzler. 2022.
\newblock \href {http://papers.nips.cc/paper\_files/paper/2022/hash/892840a6123b5ec99ebaab8be1530fba-Abstract-Conference.html} {Transformer memory as a differentiable search index}.
\newblock In \emph{NeurIPS}, pages 21831--21843.

\bibitem[{Tian et~al.(2022)Tian, Lin, Bian, Wang, and Zhao}]{cikm/Tian22TCPSRec}
Changxin Tian, Zihan Lin, Shuqing Bian, Jinpeng Wang, and Wayne~Xin Zhao. 2022.
\newblock \href {https://doi.org/10.1145/3511808.3557468} {Temporal contrastive pre-training for sequential recommendation}.
\newblock In \emph{CIKM}, pages 1925--1934.

\bibitem[{Touvron et~al.(2023)Touvron, Lavril, Izacard, Martinet, Lachaux, Lacroix, Rozi{\`{e}}re, Goyal, Hambro, Azhar, Rodriguez, Joulin, Grave, and Lample}]{TouvronLIMLLRGHARJGL23LLaMA}
Hugo Touvron, Thibaut Lavril, Gautier Izacard, Xavier Martinet, Marie{-}Anne Lachaux, Timoth{\'{e}}e Lacroix, Baptiste Rozi{\`{e}}re, Naman Goyal, Eric Hambro, Faisal Azhar, Aur{\'{e}}lien Rodriguez, Armand Joulin, Edouard Grave, and Guillaume Lample. 2023.
\newblock \href {https://doi.org/10.48550/arXiv.2302.13971} {Llama: Open and efficient foundation language models}.
\newblock \emph{CoRR}, abs/2302.13971.

\bibitem[{Wang et~al.(2020{\natexlab{a}})Wang, Louca, Hu, Cellier, Caverlee, and Hong}]{wsdm/Wang20OAR}
Jianling Wang, Raphael Louca, Diane Hu, Caitlin Cellier, James Caverlee, and Liangjie Hong. 2020{\natexlab{a}}.
\newblock \href {https://doi.org/10.1145/3336191.3371836} {Time to shop for valentine's day: Shopping occasions and sequential recommendation in e-commerce}.
\newblock In \emph{WSDM}, pages 645--653.

\bibitem[{Wang et~al.(2024{\natexlab{a}})Wang, Bao, Lin, Zhang, Li, Feng, Ng, and Chua}]{cikm24LETTER}
Wenjie Wang, Honghui Bao, Xinyu Lin, Jizhi Zhang, Yongqi Li, Fuli Feng, See{-}Kiong Ng, and Tat{-}Seng Chua. 2024{\natexlab{a}}.
\newblock \href {https://doi.org/10.1145/3627673.3679569} {Learnable item tokenization for generative recommendation}.
\newblock In \emph{CIKM}, pages 2400--2409.

\bibitem[{Wang et~al.(2024{\natexlab{b}})Wang, Cui, Fukumoto, and Suzuki}]{WangCFS24ELMRec}
Xinfeng Wang, Jin Cui, Fumiyo Fukumoto, and Yoshimi Suzuki. 2024{\natexlab{b}}.
\newblock \href {https://aclanthology.org/2024.emnlp-main.653} {Enhancing high-order interaction awareness in llm-based recommender model}.
\newblock In \emph{EMNLP}, pages 11696--11711.

\bibitem[{Wang et~al.(2022)Wang, Hou, Wang, Miao, Wu, Chen, Xia, Chi, Zhao, Liu, Xie, Sun, Deng, Zhang, and Yang}]{nips/WangHWMWCXCZL0022/NCI}
Yujing Wang, Yingyan Hou, Haonan Wang, Ziming Miao, Shibin Wu, Qi~Chen, Yuqing Xia, Chengmin Chi, Guoshuai Zhao, Zheng Liu, Xing Xie, Hao Sun, Weiwei Deng, Qi~Zhang, and Mao Yang. 2022.
\newblock \href {http://papers.nips.cc/paper\_files/paper/2022/hash/a46156bd3579c3b268108ea6aca71d13-Abstract-Conference.html} {A neural corpus indexer for document retrieval}.
\newblock In \emph{NeurIPS}.

\bibitem[{Wang et~al.(2020{\natexlab{b}})Wang, Wei, Cong, Li, Mao, and Qiu}]{sigir/Wang20GCEGNN}
Ziyang Wang, Wei Wei, Gao Cong, Xiao{-}Li Li, Xianling Mao, and Minghui Qiu. 2020{\natexlab{b}}.
\newblock \href {https://doi.org/10.1145/3397271.3401142} {Global context enhanced graph neural networks for session-based recommendation}.
\newblock In \emph{SIGIR}, pages 169--178.

\bibitem[{Xiong et~al.(2024)Xiong, Payani, Kompella, and Fekri}]{acl/Xiong24TGLLM}
Siheng Xiong, Ali Payani, Ramana Kompella, and Faramarz Fekri. 2024.
\newblock \href {https://doi.org/10.18653/v1/2024.acl-long.563} {Large language models can learn temporal reasoning}.
\newblock In \emph{ACL}, pages 10452--10470.

\bibitem[{Xu et~al.(2023)Xu, Tian, Zhang, Zhang, Wang, Zheng, Li, Tang, Zhang, Hou, Pan, Zhao, Chen, and Wen}]{sigir/23recbole1.2}
Lanling Xu, Zhen Tian, Gaowei Zhang, Junjie Zhang, Lei Wang, Bowen Zheng, Yifan Li, Jiakai Tang, Zeyu Zhang, Yupeng Hou, Xingyu Pan, Wayne~Xin Zhao, Xu~Chen, and Ji{-}Rong Wen. 2023.
\newblock \href {https://doi.org/10.1145/3539618.3591889} {Towards a more user-friendly and easy-to-use benchmark library for recommender systems}.
\newblock In \emph{SIGIR}, pages 2837--2847.

\bibitem[{Xu et~al.(2024)Xu, Hua, and Zhang}]{sigir/XuHZ24OpenP5}
Shuyuan Xu, Wenyue Hua, and Yongfeng Zhang. 2024.
\newblock \href {https://doi.org/10.1145/3626772.3657883} {Openp5: An open-source platform for developing, training, and evaluating llm-based recommender systems}.
\newblock In \emph{SIGIR}, pages 386--394.

\bibitem[{Yen et~al.(2024)Yen, Gao, and Chen}]{acl/Yen24CEPE}
Howard Yen, Tianyu Gao, and Danqi Chen. 2024.
\newblock \href {https://doi.org/10.18653/v1/2024.acl-long.142} {Long-context language modeling with parallel context encoding}.
\newblock In \emph{ACL}, pages 2588--2610.

\bibitem[{Zeghidour et~al.(2022)Zeghidour, Luebs, Omran, Skoglund, and Tagliasacchi}]{ZeghidourLOST22RQVAE}
Neil Zeghidour, Alejandro Luebs, Ahmed Omran, Jan Skoglund, and Marco Tagliasacchi. 2022.
\newblock \href {https://doi.org/10.1109/TASLP.2021.3129994} {Soundstream: An end-to-end neural audio codec}.
\newblock \emph{{IEEE} {ACM} Trans. Audio Speech Lang. Process.}, 30:495--507.

\bibitem[{Zhang et~al.(2025)Zhang, Chen, Shen, Wang, and Xiong}]{www/Zhang25HM4SR}
Shengzhe Zhang, Liyi Chen, Dazhong Shen, Chao Wang, and Hui Xiong. 2025.
\newblock \href {https://doi.org/10.1145/3696410.3714676} {Hierarchical time-aware mixture of experts for multi-modal sequential recommendation}.
\newblock In \emph{WWW}, pages 3672--3682.

\bibitem[{Zhang et~al.(2024)Zhang, Chen, Wang, Li, and Xiong}]{aaai/Zhang24TGCL4SR}
Shengzhe Zhang, Liyi Chen, Chao Wang, Shuangli Li, and Hui Xiong. 2024.
\newblock \href {https://doi.org/10.1609/aaai.v38i8.28789} {Temporal graph contrastive learning for sequential recommendation}.
\newblock In \emph{AAAI}, pages 9359--9367.

\bibitem[{Zhang et~al.(2019)Zhang, Zhao, Liu, Sheng, Xu, Wang, Liu, and Zhou}]{ZhangZLSXWLZ19FDSA}
Tingting Zhang, Pengpeng Zhao, Yanchi Liu, Victor~S. Sheng, Jiajie Xu, Deqing Wang, Guanfeng Liu, and Xiaofang Zhou. 2019.
\newblock \href {https://doi.org/10.24963/ijcai.2019/600} {Feature-level deeper self-attention network for sequential recommendation}.
\newblock In \emph{IJCAI}, pages 4320--4326.

\bibitem[{Zheng et~al.(2024)Zheng, Hou, Lu, Chen, Zhao, Chen, and Wen}]{ZhengHLCZCW24LCRec}
Bowen Zheng, Yupeng Hou, Hongyu Lu, Yu~Chen, Wayne~Xin Zhao, Ming Chen, and Ji{-}Rong Wen. 2024.
\newblock \href {https://doi.org/10.1109/ICDE60146.2024.00118} {Adapting large language models by integrating collaborative semantics for recommendation}.
\newblock In \emph{ICDE}, pages 1435--1448.

\bibitem[{Zhou et~al.(2020)Zhou, Wang, Zhao, Zhu, Wang, Zhang, Wang, and Wen}]{ZhouWZZWZWW20S3Rec}
Kun Zhou, Hui Wang, Wayne~Xin Zhao, Yutao Zhu, Sirui Wang, Fuzheng Zhang, Zhongyuan Wang, and Ji{-}Rong Wen. 2020.
\newblock \href {https://doi.org/10.1145/3340531.3411954} {S3-rec: Self-supervised learning for sequential recommendation with mutual information maximization}.
\newblock In \emph{CIKM}, pages 1893--1902.

\end{thebibliography}

\appendix

\newpage 
\newpage

\appendix

\section{Additional Related Work}\label{sec:app_relatedwork}

\subsection{Sequential Recommendation}\label{sec:app_relatedwork_sr}
The goal of sequential recommendation is to predict the following items that users may be interested in based on their behavior sequences. Early works focus on various neural-based encoders, such as convolutional neural networks~\cite{TangW18caser}, gated recurrent units~\cite{HidasiKBT15GRU4Rec}, and Transformers~\cite{KangM18SASRec, SunLWPLOJ19BERT4Rec, kdd/MaKL19HGN}.
Recent approaches incorporate item textual attributes, employing separate self-attention mechanisms for item and feature information~\cite{ZhangZLSXWLZ19FDSA}, while leveraging self-supervised objectives to learn item-attribute correlations~\cite{ZhouWZZWZWW20S3Rec}.
However, these models are limited in fully utilizing the reasoning power of LLMs and textual semantics, unlike generative recommendation approaches.

\subsection{LLM-based Recommendation}\label{sec:app_relatedwork_llmrec}
Recent studies~\cite{BaoZZWF023TALLRec, LCZZX23E4SRec, sigir/Liao24LLaRA, KimKC0YP24A-LLMRec} employ LLMs directly as re-rankers, where the model is prompted with a subset of item candidates (typically 20 items, including one ground-truth item) to recommend items likely to be preferred by users. These approaches utilize the rich knowledge and reasoning capabilities of LLMs to enhance recommendation quality. 
Meanwhile, some works~\cite{RenWXSCWY024RLMRec, Liu0WLLM-ESR, WZ00025LLMEmb, SZZCWC25AlphaRec} only extract LLM knowledge to initialize or enhance traditional recommendation models, avoiding the costly LLM fine-tuning.
Unlike these approaches, our work focuses on direct item ID generation, performing full ranking across the entire item space rather than re-ranking from sampled candidates.

\begin{table}[t]  \small
\begin{center}
\centering
\scalebox{0.9}{
    \begin{tabular}{c|cccc}
    \toprule
    Dataset & Beauty & Toys & Sports & Yelp  \\
    \midrule
    \#Users & 22,363 & 19,412 & 35,598 & 30,431 \\
    \#Items & 12,101 & 11,924 & 18,357 & 20,033 \\
    \#Inter. & 198,502 & 167,597 & 296,337 & 316,354 \\
    Density & 0.0734\% & 0.0724\% & 0.0453\% & 0.0519\% \\
    Avg. Length & 8.9 & 8.6 & 8.3 & 10.4 \\
    Avg. Interval & 69.6d & 86.0d & 74.1d & 18.6d \\
    \bottomrule
    \end{tabular}
}
\caption{Statistics of four benchmark datasets.}
\label{tab:statistics}
\vspace{-2mm}
\end{center}
\end{table}
\begin{table}[t] \small
\centering
\vspace{-2mm}
\begin{tabular}{c|ccc}
\toprule
Dataset & Short & Middle & Long \\ \midrule
Beauty & 9,719 & 5,052 & 7,592 \\
Toys & 9,518 & 3,323 & 6,571 \\ 
\bottomrule
\end{tabular}
\caption{The number of users of each test subset in Figure~\ref{fig:exp_time_interval_group}, categorized by time interval between the most recent interaction and the target item.}
\label{tab:statistics_interval}
\vspace{-2mm}
\end{table}

\section{Additional Experimental Setup}\label{sec:app_expsetup}
\subsection{Datasets}\label{sec:app_expsetup_data}
Following the previous works~\cite{Tan24IDGenRec, WangCFS24ELMRec, Geng0FGZ22P5, HuaXGZ23P5Howtoindex}, we use the Amazon Review dataset containing product reviews and item metadata from 1996 to 2014. We also use the Yelp dataset with business reviews from 2019 to 2020. Table~\ref{tab:statistics} presents statistics of preprocessed datasets. We further provide the number of users for each subset in Figure~\ref{fig:exp_time_interval_group}.

\subsection{Baselines}\label{sec:app_expsetup_baseline}
We adopt six traditional models, four temporal models, and eight generative models for baselines.

\begin{itemize}[leftmargin=*,topsep=0pt,itemsep=-1ex,partopsep=1ex,parsep=1ex]

\item \textbf{GRU4Rec}~\cite{HidasiKBT15GRU4Rec} encodes sequential user behavior using Gated Recurrent Units.
\item \textbf{HGN}~\cite{kdd/MaKL19HGN} models long- and short-term interests with a hierarchical gating network.
\item \textbf{SASRec}~\cite{KangM18SASRec} leverages uni-directional Transformers to represent users based on their most recent interaction.
\item \textbf{BERT4Rec}~\cite{SunLWPLOJ19BERT4Rec} employs bi-directional self-attention for masked item prediction tasks.
\item \textbf{FDSA}~\cite{ZhangZLSXWLZ19FDSA} separately models feature-level and item-level self-attention.
\item \textbf{S$^3$Rec}~\cite{ZhouWZZWZWW20S3Rec} enhances representation learning with self-supervised auxiliary tasks.
\item \textbf{TiSASRec}~\cite{LiWM20TISASREC} introduces relative time interval embeddings as keys and values in self-attention mechanisms.
\item \textbf{TiCoSeRec}~\cite{aaai/Dang23TiCoSeRec} improves contrastive learning by augmenting sequences with controlled time interval distributions.
\item \textbf{HM4SR}~\cite{www/Zhang25HM4SR} employs a mixture of experts architecture to integrate temporal patterns with multi-modal (ID, text) representations.
\item \textbf{HORAE}~\cite{TOIS/Hu25HORAE} enhances multi-interest learning with temporal dynamics.
\item \textbf{P5-SID}~\cite{HuaXGZ23P5Howtoindex} assigns numeric IDs sequentially based on the item appearance. 
\item \textbf{P5-CID}~\cite{HuaXGZ23P5Howtoindex} clusters items based on co-occurrences to generate numeric IDs.
\item \textbf{P5-SemID}~\cite{HuaXGZ23P5Howtoindex} assigns numeric IDs using item metadata like categories.
\item \textbf{TIGER}~\cite{RajputMSKVHHT0S23TIGER} introduces codebook IDs generated through RQ-VAE. 
\item \textbf{IDGenRec}~\cite{Tan24IDGenRec} generates textual IDs with a generator based on item metadata.
\item \textbf{ELMRec}~\cite{ChenLLMX22ICLRec} adopts high-order relationships using soft prompts and re-ranking strategies with numeric IDs.
\item \textbf{LETTER}~\cite{cikm24LETTER} integrates hierarchical semantics, collaborative signals, and diversity with RQ-VAE IDs.
\item \textbf{LC-Rec}~\cite{ZhengHLCZCW24LCRec} combines RQ-VAE IDs with multi-task learning to integrate language and collaborative semantics. 
\item \textbf{GRAM}~\cite{acl25GRAM} translates implicit item relationships and employs multi-granular late fusion to integrate rich item semantics.
\end{itemize}

\begin{table}[t]\footnotesize
\centering
\vspace{-3mm}
\setlength{\tabcolsep}{4.5pt}
\renewcommand{\arraystretch}{0.8}
\begin{tabular}{c|cccc}
\toprule
Hyperparameters & Beauty & Toys & Sports & Yelp \\
\midrule
$\epsilon$ & 0.01 & 0.01 & 0.001 & 0.01 \\
$k$ & 1 & 1 & 1 & 1 \\
$L$ & 5 & 2 & 3 & 3 \\
$\lambda$ & 0.3 & 0.4 & 0.1 & 0.2 \\
$N$ & 7 & 30 & 30 & 30 \\
$\tau$ & 128 & 1024 & 128 & 256 \\
$c$ & 0.8 & 0.8 & 0.9 & 0.8 \\
\bottomrule
\end{tabular}
\caption{Final hyperparameters for \ours.}
\label{tab:final_hyper}
\vspace{-5mm}
\end{table}

\subsection{Additional Implementation Details}\label{sec:app_expsetup_detail}

We conducted all experiments with 2 NVIDIA RTX A6000, 512 GB memory, and 2 AMD EPYC 74F3. 

\subsubsection{Details for \ours}\label{sec:app_expsetup_detail_grut}

We implemented \ours~on OpenP5~\cite{sigir/XuHZ24OpenP5}. We tuned $\epsilon$ in $\{10^{-3},10^{-2}, 10^{-1}, 10^{0}\}$, $k$ in $\{1, 3, 5, 10\}$, $L$ in $\{1, 2, 3, 4, 5, 7\}$, $\lambda$ in the range of $[0, 1]$ with step size 0.1, $N$ in $\{ 7, 30, 180, 360\}$, $\tau$ in the range of $[2^{7}, 2^{11}]$ with exponentially increasing steps in powers of 2, $c$ in $[0.8,1]$. Due to computational constraints, hyperparameters were tuned sequentially. We first optimize $\epsilon$, followed by $k$, $L$, $\tau$, $c$, $\lambda$, and finally $N$. The final hyperparameters are in Table~\ref{tab:final_hyper}. We sort the user history in $C_u$ and $C_v$ in reverse order to prevent recent items from being truncated following the existing work~\cite{LiWLFSSM23Recformer}. For hyperparameter sensitivity analysis (Figure~\ref{fig:exp_hyper_neighbor},~\ref{fig:exp_hyper_window}, and~\ref{fig:exp_hyper_epsilon}), we measured performance without trend-aware inference to ensure a more accurate analysis, \ie, $\lambda$ = 0. For calculating the trend score in Eq.~\eqref{eq:trend_score}, the recommendation day itself was excluded, \ie, from day $t_{|s_u|+1}-N-1$ to day $t_{|s_u|+1}-1$.

The keywords are extracted from each item's textual metadata and assigned as textual IDs following the existing work~\cite{acl25GRAM}. Rather than learning an ID generator during training~\cite{Tan24IDGenRec}, we precompute TF-IDF scores~\cite{Jones2004/TFIDF} over the metadata before training. We then select the highest-scoring terms and assign them as IDs. To maintain consistency with a backbone LLM, the T5 tokenizer~\cite{JMLR/Raffel2020/T5} is adopted. For the Amazon Beauty, Toys, and Sports dataset, we concatenate each item's title, brand, category, and description. The name, city, and category fields are used for the Yelp dataset. For the Toys dataset, examples of item IDs include `\textit{musical-piano-concert-keyboard-displays}', `\textit{dinosaur-safari-dragon-knight-headed}', and `\textit{doll-loving-bedroom-mirrored-comfy}'.

\subsubsection{Details for Baselines}\label{sec:app_expsetup_detail_baseline}

For traditional and temporal recommendation models, we conducted all experiments on the open-source RecBole library~\cite{sigir/23recbole1.2}\footnote{\url{https://recbole.io/}}. We thoroughly tuned each hyperparameter following guidance from the original papers. The models were optimized using the Adam optimizer with a learning rate of 0.001, a batch size of 256, and an embedding dimension of 64. The training was stopped when the validation NDCG@10 showed no improvement for 10 consecutive epochs. For HM4SR~\cite{www/Zhang25HM4SR}, we utilized only ID and text embeddings, without image embeddings, to ensure a fair comparison. While HORAE~\cite{TOIS/Hu25HORAE} used Amazon 2018 datasets, we pre-trained the model with the corresponding Amazon 2014 datasets (Food, CDs, Kindle, Movies, and Home) for consistency with our experimental setup, then fine-tuned the pre-trained model on Beauty, Toys, Sports, and Yelp datasets, respectively.

For all generative baselines, we follow the official code if publicly available, \eg, 
P5-variants~\cite{HuaXGZ23P5Howtoindex}\footnote{\url{https://github.com/Wenyueh/LLM-RecSys-ID}}, 
IDGenRec~\cite{Tan24IDGenRec}\footnote{\url{https://github.com/agiresearch/IDGenRec}}, 
ELMRec~\cite{WangCFS24ELMRec}\footnote{\url{https://github.com/WangXFng/ELMRec}}, 
LC-Rec~\cite{ZhengHLCZCW24LCRec}\footnote{\url{https://github.com/RUCAIBox/LC-Rec}}, 
LETTER~\cite{cikm24LETTER}\footnote{\url{https://github.com/HonghuiBao2000/LETTER}}, and GRAM~\cite{acl25GRAM}\footnote{\url{https://github.com/skleee/GRAM}}. For TIGER~\cite{RajputMSKVHHT0S23TIGER}, we implemented the model based on the details in the paper since the official code was not publicly available. We used the Sentence-T5~\cite{acl/NiACMHCY22/SentenceT5} for semantic embeddings with a hidden dimension size of 768. The vocabulary size was set to 1024 (256$\times$4). We used T5-small~\cite{JMLR/Raffel2020/T5} for P5, IDGenRec, and ELMRec, following the official codebase. We instantiate LETTER on TIGER. 

For ELMRec, when applying to the Yelp dataset, which is not included in the original paper, we excluded the explanation generation task due to insufficient textual metadata. Additionally, we did not apply the `reranking approach' proposed in ELMRec for the Yelp dataset since items within a user sequence can reappear as target items. For all other implementation details, including hyperparameter search ranges, we thoroughly followed the specifications described in the ELMRec manuscript.

For LC-Rec, we fully fine-tuned LLaMA-7B~\cite{TouvronLIMLLRGHARJGL23LLaMA}, adhering to the authors’ guidelines with some modifications for the Amazon 2014 dataset. In the asymmetric item prediction task, we set the number of training samples based on the interactions for each dataset, \eg, 20K, 15K, 25K, and 25K for the Beauty, Toys, Sports, and Yelp datasets, respectively. For the personalized preference inference task, we used \texttt{gpt-4o-mini-2024-07-18} to infer user preferences on Amazon datasets and omitted this task on Yelp due to insufficient textual metadata.

\subsubsection{Modifications to Preprocessing of ELMRec and IDGenRec}\label{sec:app_expsetup_detail_mod}

For ELMRec, we adopted the P5-SID used in the official code with modifications to prevent data leakage. Following recent works~\cite{HuaXGZ23P5Howtoindex, sigir/XuHZ24OpenP5, RajputMSKVHHT0S23TIGER}, we excluded validation and test items while assigning numeric IDs. The original P5 methodology assigned consecutive IDs to items based on their appearance order within each user sequence. For instance, a user sequence is represented as \texttt{[8921, 8922, ..., 8927]}, where \texttt{8927} becomes the test item in the leave-one-out evaluation. Since P5 uses the SentencePiece tokenizer~\cite{Sennrich16aSentencePiece}, test items potentially share subwords with training items. It creates unintended correlations that implicitly lead to information leakage during inference, as already identified in previous works~\cite{RajputMSKVHHT0S23TIGER, kdd24TransRec}\footnote{Please refer to Appendix D of \citet{RajputMSKVHHT0S23TIGER} and Appendix A.6 of \citet{kdd24TransRec} for details.}.

For IDGenRec, we excluded user IDs from input prompts, following guidance from the original authors\footnote{\url{https://github.com/agiresearch/IDGenRec/issues/1}}. This explains the differences in performance in Table~\ref{tab:exp_overall} compared to the original paper. Initially, IDGenRec uses both item IDs and a user ID. The user ID is created by concatenating all sequence items and processing them through the ID generator. For example, with an item sequence $i_1 \rightarrow i_2 \rightarrow i_3 \rightarrow i_4$, information from all items is used. However, this approach creates a potential data leakage issue in leave-one-out evaluation settings, as the user ID contains information about the test item $i_4$. To address this concern, we removed user IDs from our implementation.

\subsection{Examples of Input Prompts for Table~\ref{tab:exp_time_info}}\label{sec:app_expsetup_ablation_input}
We present six types of user-level temporal context $C_u$ shown in Table~\ref{tab:exp_time_info}. 

\begin{tcolorbox}
\textbf{None}: \\
What would the user purchase after \\
$\Tilde{i}_1$, 
$\Tilde{i}_2$, 
$\cdots$, 
$\Tilde{i}_{|s_u|}$ ?
\end{tcolorbox}

\begin{tcolorbox}
\textbf{Absolute}: \\
What would the user purchase after \\
$\Tilde{i}_1$ ($t_1$), 
$\Tilde{i}_2$ ($t_2$), 
$\cdots$, 
$\Tilde{i}_{|s_u|}$ ($t_{|s_u|}$) ?
\end{tcolorbox}

\begin{tcolorbox}
\textbf{Relative}: \\
What would the user purchase after \\
$\Tilde{i}_1$ (after $t_2-t_1$), 
$\Tilde{i}_2$ (after $t_3-t_2$), \\
$\cdots$, $\Tilde{i}_{|s_u|}$ ?
\end{tcolorbox}

\begin{tcolorbox}
\textbf{Target-relative}: \\
What would the user purchase after \\
$\Tilde{i}_1$ ($t_{|s_u|+1} - t_1$ ago), 
$\Tilde{i}_2$ ($t_{|s_u|+1} - t_2$ ago), \\
$\cdots$, 
$\Tilde{i}_{|s_u|}$ ($t_{|s_u|+1} - t_{|s_u|}$ ago) ?
\end{tcolorbox}

\begin{tcolorbox}
\textbf{Relative + Absolute}: \\
What would the user purchase after \\
$\Tilde{i}_1$ ($t_1$, after $t_2-t_1$), 
$\Tilde{i}_2$ ($t_2$, after $t_2-t_2$), \\
$\cdots$, 
$\Tilde{i}_{|s_u|}$ ($t_{|s_u|}$) ?
\end{tcolorbox}

\begin{tcolorbox}
\textbf{Target-relative + Absolute}: \\
The current date is $t_{|s_u|+1}$. \\ What would the user purchase after \\
$\tilde{i}_1$ ($t_1$, $\Delta t_1$ ago), 
$\tilde{i}_2$ ($t_2$, $\Delta t_2$ ago), 
$\cdots$, \\
$\tilde{i}_{|s_u|}$ ($t_{|s_u|}$, $\Delta t_{|s_u|}$ ago) ?
\end{tcolorbox}

\begin{figure}[t]\small
\centering
\includegraphics[width=1.0\linewidth]{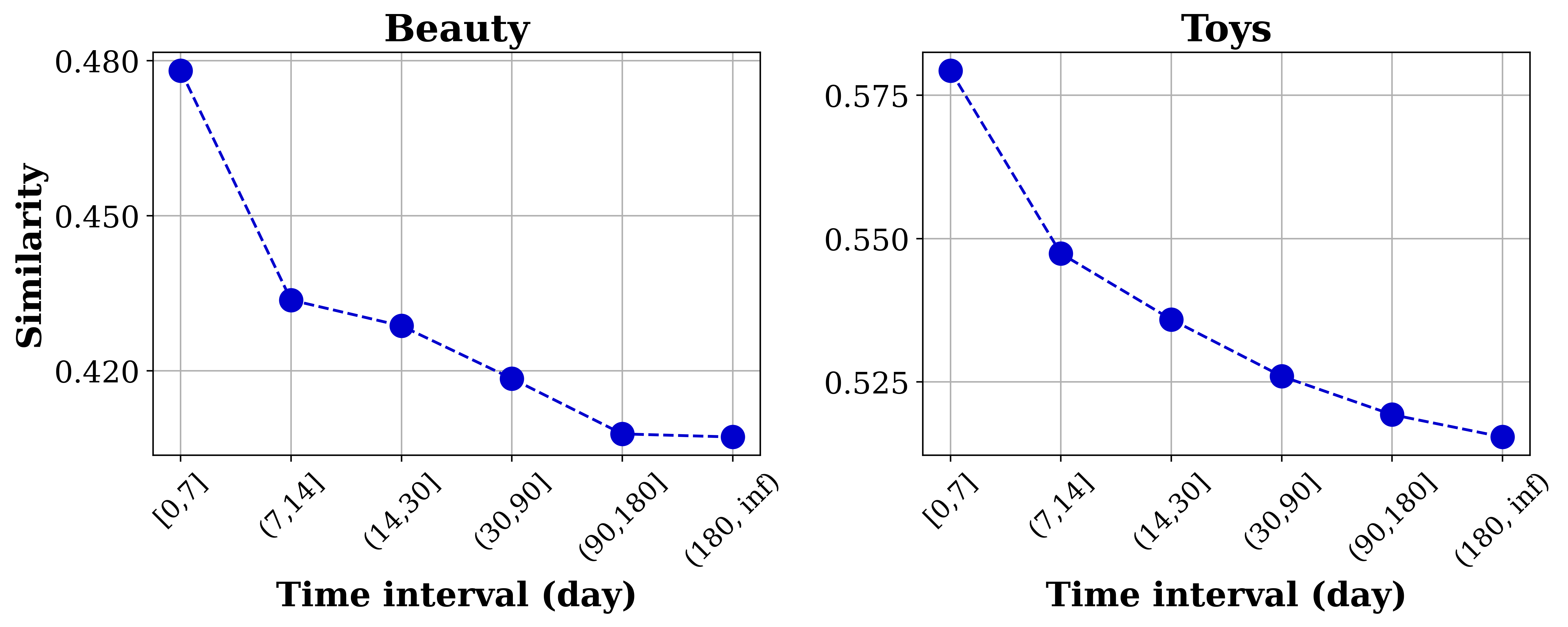}
\includegraphics[width=1.0\linewidth]{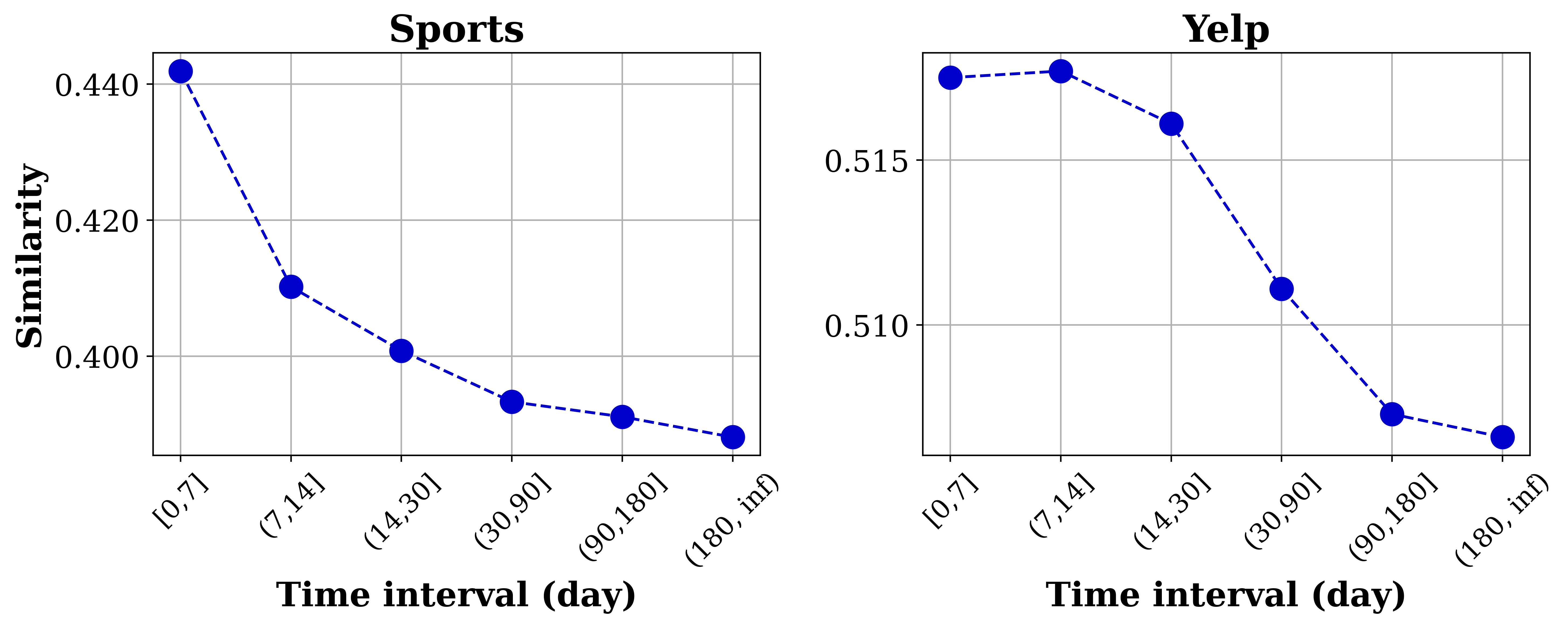}
\caption{Similarity of item pairs by time interval groups. The x-axis is the time interval between two consecutive items, and the y-axis is the semantic similarity of items.}\label{fig:fig_time_interval_sim}
\end{figure}

\begin{table}[t]\footnotesize\centering
\vspace{-1.5mm}
\setlength{\tabcolsep}{3.8pt}
\renewcommand{\arraystretch}{1}
\begin{tabular}{c|c|cc|cc}
\toprule
\multirow{2}{*}{ID} & \multirow{2}{*}{Temporal}    & \multicolumn{2}{c|}{Beauty} & \multicolumn{2}{c}{Toys} \\
&        & R@5      & N@5         & R@5         & N@5       \\
\midrule
\multirow{2}{*}{Ours}        & \cmark & \textbf{0.0740} & \textbf{0.0511} & \textbf{0.0769} & \textbf{0.0531} \\
                             & \xmark & 0.0582 & 0.0404 & 0.0558 & 0.0392 \\
\midrule
\multirow{2}{*}{IDGenRec} & \cmark & \textbf{0.0733} & \textbf{0.0516} & \textbf{0.0659} & \textbf{0.0454} \\
                             & \xmark & 0.0533 & 0.0374 & 0.0487 & 0.0329 \\
\midrule
\multirow{2}{*}{Title ID} & \cmark & \textbf{0.0572} & \textbf{0.0394} & \textbf{0.0558} & \textbf{0.0400} \\
                             & \xmark & 0.0411 & 0.0293 & 0.0444 & 0.0314 \\
\bottomrule
\end{tabular}
\caption{Performance of \ours~over various IDs. }\label{tab:various_ids}
\vspace{-4mm}

\end{table}

\begin{table}[t]\footnotesize\centering
\vspace{-1.5mm}
\setlength{\tabcolsep}{4.0pt}
\renewcommand{\arraystretch}{1}
\begin{tabular}{c|c|cc|cc}
\toprule
\multirow{2}{*}{Model} & \multirow{2}{*}{Trend}    & \multicolumn{2}{c|}{Beauty} & \multicolumn{2}{c}{Toys} \\
&        & R@5      & N@5         & R@5         & N@5       \\
\midrule
\multirow{2}{*}{IDGenRec}        & \cmark & \textbf{0.0480} & \textbf{0.0332} & \textbf{0.0480} & \textbf{0.0328} \\
                             & \xmark & 0.0463 & 0.0328 & 0.0462 & 0.0323 \\
\midrule
\multirow{2}{*}{LETTER} & \cmark & \textbf{0.0373} & \textbf{0.0250} & \textbf{0.0324} & \textbf{0.0211} \\
                             & \xmark & 0.0364 & 0.0243 & 0.0309 & 0.0202 \\
\midrule
\multirow{2}{*}{LC-Rec} & \cmark & \textbf{0.0521} & \textbf{0.0365} & \textbf{0.0574} & \textbf{0.0399} \\
                             & \xmark & 0.0503 & 0.0352 & 0.0543 & 0.0385 \\
\bottomrule
\end{tabular}
\caption{Effectiveness of trend-aware inference when applying to existing generative recommenders.}\label{tab:trend_general}
\end{table}

\begin{table}[t]\footnotesize\centering
\begin{tabular}{c|cccc}
\toprule
Phase & Beauty & Toys & Sports & Yelp \\ \midrule
Offline & 2.5 & 2.6 & 3.2 & 3.6 \\
Online & 1.7 & 1.6 & 2.1 & 2.2 \\ \midrule
Total & 4.2 & 4.1 & 5.3 & 5.8 \\ \bottomrule
\end{tabular}
\caption{Latency (milliseconds per user) for the offline trend score computation and online score aggregation. }\label{tab:latency}
\vspace{-2mm}
\end{table}

\begin{figure}[t]\small
\centering
\includegraphics[width=1.0\linewidth]{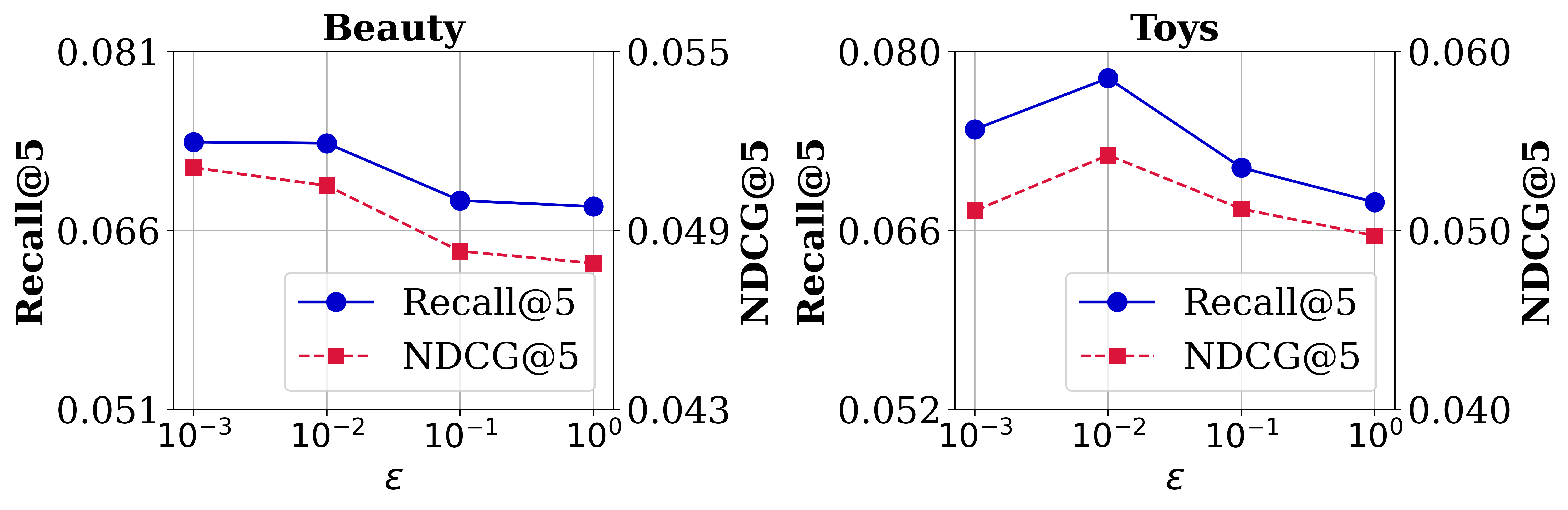}
\includegraphics[width=1.0\linewidth]{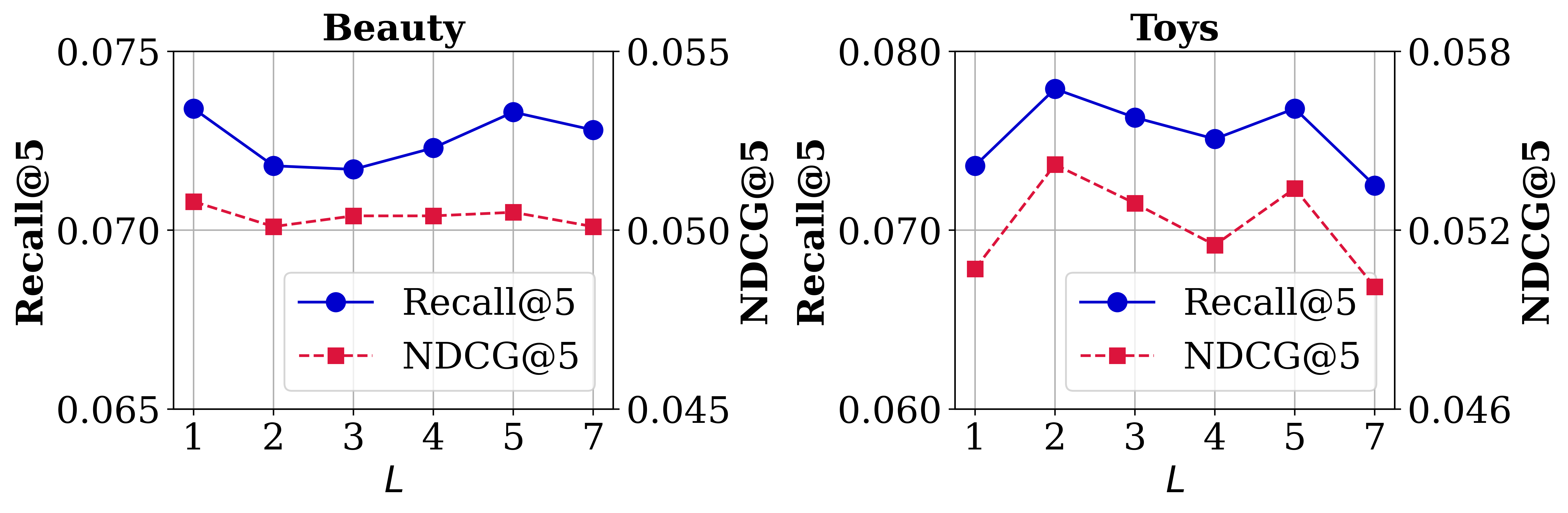} 
\vspace{-6.5mm}
\caption{Performance of \ours~over varying (i) $\epsilon$ that controls the influence of item-level transition patterns and (ii) the number of the most recent items $L$ in $C_v$.}\label{fig:exp_hyper_epsilon}
\vspace{-3.5mm}
\end{figure}

\section{Additional Experimental Results}\label{sec:app_expresult}

\subsection{Preference Shifts over Time Interval}\label{sec:app_expresult_shift}
We examined whether user preferences evolve over time by analyzing item similarity across different time intervals. Figure~\ref{fig:fig_time_interval_sim} shows text similarity between consecutive items grouped by time intervals. For calculating similarity, we generated text embeddings using NVEmbed~\cite{abs-2405-17428/NVEmbed} from item metadata\footnote{For Amazon datasets, we used title, brand, and categories. We used name, city, and categories for the Yelp dataset.}. Each consecutive item pair from user sequences is grouped by time intervals of interaction, \eg, intervals of 8 days fall into $(7, 14]$, and interactions of the same day belong to $[0,7]$. The results clearly show decreasing similarity between consecutive items as time intervals increase across all datasets. It suggests that user preferences shift more significantly over longer time intervals. Despite these challenges, our model demonstrates superior performance, especially in scenarios with long time gaps, as demonstrated in Figure~\ref{fig:exp_time_interval_group}.

\subsection{Effect of ID Variants}\label{sec:app_expresult_id}
Table~\ref{tab:various_ids} demonstrates the effectiveness of \ours~across different ID variants. When replacing our IDs with those from prior work~\cite{Tan24IDGenRec} or titles\footnote{We appended additional digits for duplicated titles to ensure uniqueness.}, \ours~consistently improves performance, enhancing R@5 and N@5 by 27.1\%--39.2\% and 26.6\%--38.0\%, respectively. It implies the robustness of our temporal integration approach regardless of ID schemes.

\subsection{Generalizability of Trend-aware Inference}\label{sec:app_expresult_trend}
Table~\ref{tab:trend_general} illustrates the effect of trend-aware inference when applying to existing generative recommendation baselines, \eg, IDGenRec, LETTER, and LC-Rec. Notably, all baselines consistently show performance improvements, achieving average gains of 4.0\% in R@5 and 2.9\% in N@5, respectively. This confirms that our trend-aware inference effectively enhances recommendation performance regardless of the underlying architecture.

\subsection{Analysis of Computational Overhead}\label{sec:app_expresult_latency}
Table~\ref{tab:latency} shows the computational overhead of trend-aware inference by measuring the latency of its offline trend score computation and online score aggregation phases. The additional online latency is 1.6–2.2 ms per user, which is marginal compared to the beam search time (0.1–0.3s per user). It confirms that trend-aware inference is practically feasible with minimal overhead.

\subsection{Hyperparameter Sensitivity}\label{sec:app_expresult_hyper}
Figures \ref{fig:exp_hyper_epsilon} shows the performance of \ours~depending on $\epsilon$, which controls the influence of $C_v$, and the number of the most recent items in $C_v$, denoted as $L$. For $\epsilon$, the optimal values for Beauty and Toys are 0.001 and 0.01, respectively. This suggests that a large $\epsilon$ makes the model excessively focus on item transition patterns, neglecting user-specific signals. Meanwhile, the optimal values of $L$ for Beauty and Toys are 1 and 2, respectively. It highlights how the dataset characteristics directly influence the optimal hyperparameters.

\section{Additional Case Study}\label{sec:app_case}

\subsection{Effect of Inference Timestamp Shift}\label{sec:app_case_shift}
Table~\ref{fig:case_study_shift} presents the recommendation results of \ours~for the same user, evaluated at different inference timestamps ($\Delta t_{|s_u|+1}$) in the user-level temporal context $C_u$. When the inference occurs shortly after the user's last interaction, \ours~emphasizes short-term interests, recommending products related to the most recent purchase, \eg, `\textit{Sofia}'. In contrast, when the inference timestamp is distant from the last interaction, the model recommends items reflecting long-term interests, \eg, `\textit{RC helicopters}', which had been frequently purchased in the past. These results demonstrate \ours's ability to adapt recommendations based on inference timestamp, unlike existing generative recommendation models that produce identical predictions regardless of when inference occurs.

\newcolumntype{C}[1]{>{\centering\arraybackslash}m{#1\columnwidth}}

\begin{table}[t]\scriptsize
\centering
\setlength{\tabcolsep}{2pt}
\renewcommand{\arraystretch}{0.4}

\begin{tabular}{C{0.125}|*{5}{C{0.15}}}
\toprule
\multicolumn{6}{c}{\textbf{User sequence (\valfont{ASIN: A2N8D20LSUU85O})}} \\ \midrule

\textbf{Image} &
\cellimg{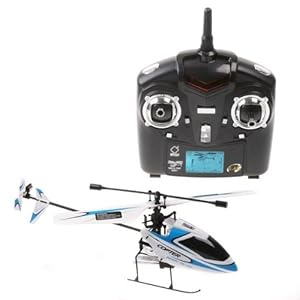} &
\cellimg{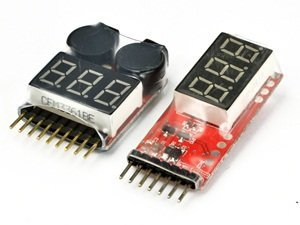} &
\cellimg{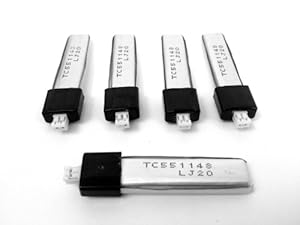} &
\cellimg{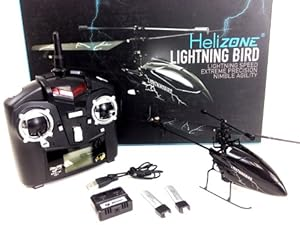} &
\cellimg{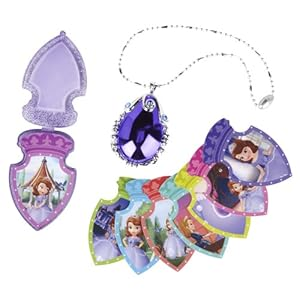} \\ \midrule

\textbf{Name}     & \valfont{\shortstack{WL V911\\RC Helicopter}}
         & \valfont{\shortstack{Battery\\Checker}} 
         & \valfont{\shortstack{WL V911\\Battery 5-Pack}}
         & \valfont{\shortstack{Helizone\\Edition}} 
         & \valfont{\shortstack{Sofia\\Amulet}} \\ \midrule
\textbf{Category} & \valfont{\shortstack{RC\\Helicopters}}
         & \valfont{\shortstack{Battery\\Chargers}}
         & \valfont{\shortstack{Vehicle\\Batteries}}
         & \valfont{\shortstack{RC\\Propellers}}
         & \valfont{\shortstack{Pretend\\Play}} \\ \midrule
\textbf{Time}     & \valfont{2013-02-18} & \valfont{2013-02-18} & \valfont{2013-02-18} & \valfont{2013-05-03} & \valfont{2013-12-30} \\ \bottomrule


\multicolumn{6}{c}{} \\[0.4em] 
\rowcolor{gray!20} 
\multicolumn{6}{c}{\textbf{GRUT Top 5 prediction at 2013-12-30}} \\ \midrule

\textbf{Ranking} & 1 & 2 & 3 & 4 & 5 \\ \midrule

\textbf{Image} &
\cellimg{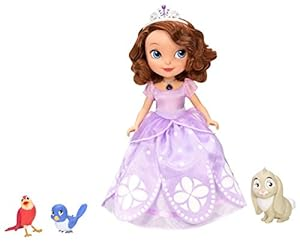} &
\cellimg{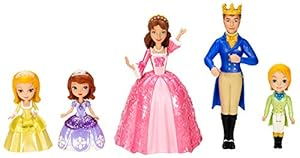} &
\cellimg{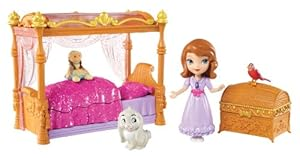} &
\cellimg{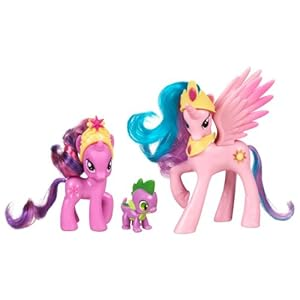} &
\cellimg{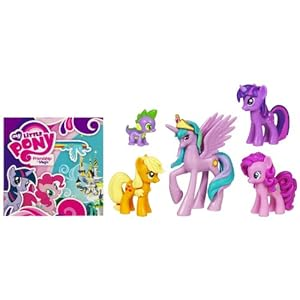} \\ \midrule

\textbf{Name}     & \valfont{\shortstack{Sofia\\Animals}} 
         & \valfont{\shortstack{Sofia\\Royal Family}}
         & \valfont{\shortstack{Sofia\\Royal Bed}}
         & \valfont{\shortstack{Magic\\Castle Friends}}
         & \valfont{\shortstack{Magic\\Gift Set}} \\ \midrule
\textbf{Category} & \valfont{\shortstack{Dolls\&\\Playsets}}
         & \valfont{Dolls}
         & \valfont{Playsets}
         & \valfont{\shortstack{Action Figs.\\\& Playsets}}
         & \valfont{Playsets} \\ \bottomrule


\multicolumn{6}{c}{} \\[0.4em] 
\rowcolor{gray!20} 
\multicolumn{6}{c}{\textbf{GRUT Top 5 prediction at 2014-06-30}} \\ \midrule

\textbf{Ranking} & 1 & 2 & 3 & 4 & 5 \\ \midrule

\textbf{Image} &

\cellimg{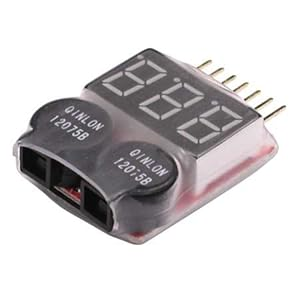} &
\cellimg{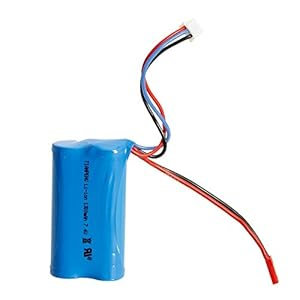} &
\cellimg{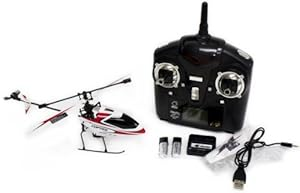} &
\cellimg{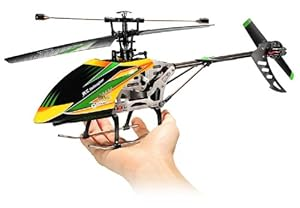} &
\cellimg{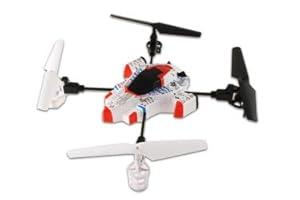} \\ \midrule

\textbf{Name}    & \valfont{\shortstack{Voltage\\Checker}} 
                 & \valfont{\shortstack{Double Horse\\9053 Gyro}} 
                 & \valfont{\shortstack{WL V911\\Red V2}} 
                 & \valfont{\shortstack{WL V912\\Gyro RTF}} 
                 & \valfont{\shortstack{Syma\\Quad Copter}} \\ \midrule
\textbf{Category} & \valfont{\shortstack{Vehicle\\Batteries}}
                 & \valfont{\shortstack{Vehicle\\Batteries}}
                 & \valfont{RC Helicopters}
                 & \valfont{RC Helicopters}
                 & \valfont{RC Helicopters} \\ \bottomrule

\end{tabular}

\caption{\ours’s top-5 predictions on the Toys dataset at different inference timestamps. The five most recent items in the sequence are shown for simplicity. }\label{fig:case_study_shift}
\vspace{-3mm}
\end{table}

\newcolumntype{C}[1]{>{\centering\arraybackslash}m{#1\columnwidth}}

\begin{table}[t]\scriptsize
\centering
\setlength{\tabcolsep}{2pt}
\renewcommand{\arraystretch}{0.4}

\begin{tabular}{C{0.125}|*{5}{C{0.15}}}

\toprule
\multicolumn{6}{c}{\textbf{User sequence (\valfont{ASIN: A2V65NBADV4HY4})}} \\ \midrule

\textbf{Image} &
\cellimg{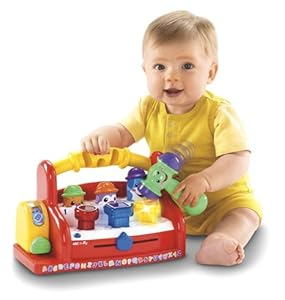} &
\cellimg{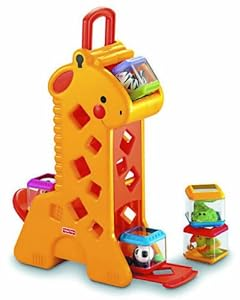} &
\cellimg{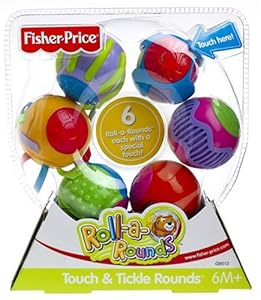} &
\cellimg{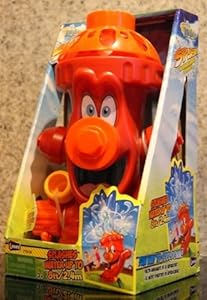} &
\cellimg{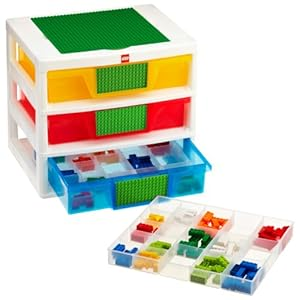} \\ \midrule

\textbf{Name}     & \valfont{\shortstack{Learning\\Toolbench}} 
                 & \valfont{\shortstack{Peek-a-Blocks\\Giraffe}} 
                 & \valfont{\shortstack{Touch \&\\Tickle Rounds}} 
                 & \valfont{\shortstack{Garden Hose\\Sprinkler}} 
                 & \valfont{\shortstack{LEGO\\Sorting System}} \\ \midrule 

\textbf{Category} & \valfont{\shortstack{Learning\\Toys}} 
                 & \valfont{\shortstack{Baby\\Toys}} 
                 & \valfont{\shortstack{Gag\\Toys}} 
                 & \valfont{\shortstack{Outdoor\\Toys}} 
                 & \valfont{\shortstack{Building\\Toys}} \\ \midrule

\textbf{Time}     & \valfont{2005-10-31} & \valfont{2006-08-03} & \valfont{2006-08-03} & \valfont{2013-11-19} & \valfont{2014-01-01} \\ \bottomrule

\multicolumn{6}{c}{} \\[0.4em] 
\rowcolor{gray!20} 
\multicolumn{6}{c}{\textbf{GRUT Top 5 prediction at 2014-01-01 ($\lambda=0.0$)}} \\ \midrule

\textbf{Ranking} & 1 & 2 & 3 & 4 & 5 \\ \midrule

\textbf{Image} &
\cellimg{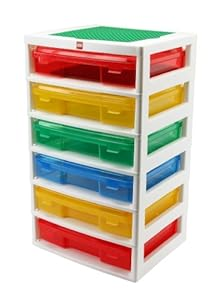} &
\cellimg{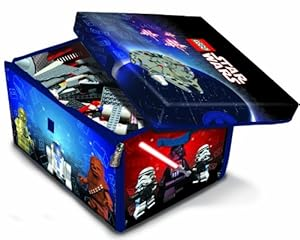} &
\cellimg{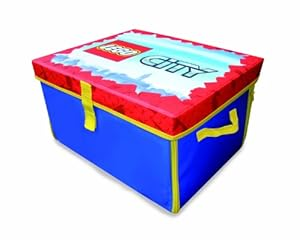} &
\cellimg{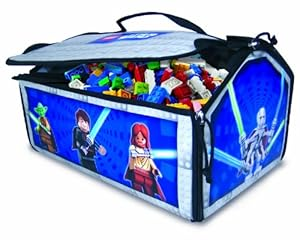} &
\cellimg{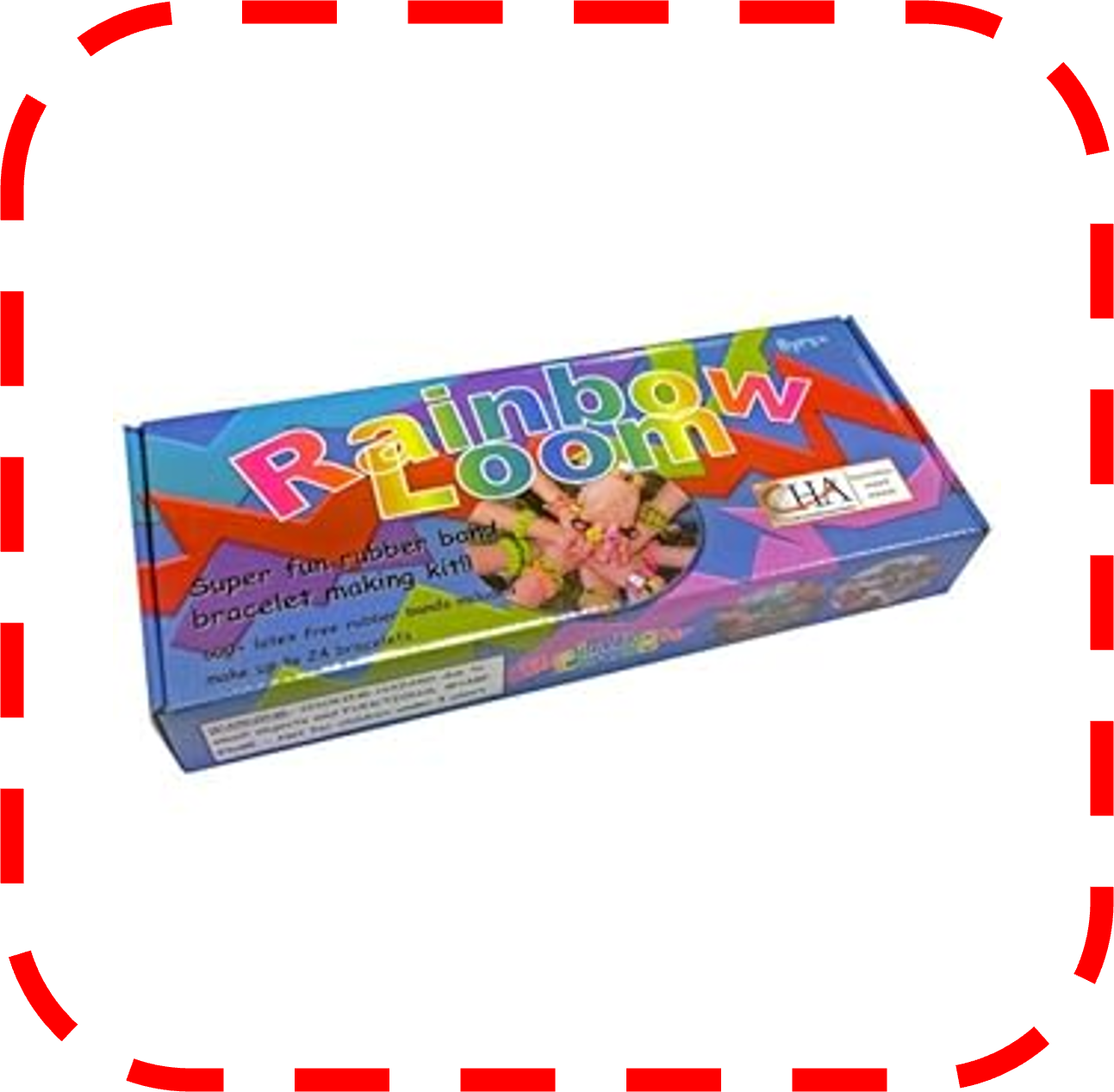} 
\\ \midrule

\textbf{Name}     & \valfont{\shortstack{LEGO 6-Case\\Storage Unit}} 
                 & \valfont{\shortstack{Star Wars\\Box}} 
                 & \valfont{\shortstack{LEGO City\\Box}} 
                 & \valfont{\shortstack{Star Wars\\Battle Bridge}} 
                 & \valfont{\shortstack{Rainbow\\Loom}} \\ \midrule

\textbf{Category} & \valfont{\shortstack{Building\\Toys}} 
                 & \valfont{\shortstack{Vehicle\\Playsets}} 
                 & \valfont{\shortstack{Vehicle\\Playsets}} 
                 & \valfont{\shortstack{Toys \&\\Games}} 
                 & \valfont{\shortstack{Toys \&\\Games}} \\ \midrule

\textbf{\(s_{\text{trend}}\)} 
                & \valfont{0.1361}
                & \valfont{0.0206} 
                & \valfont{0.0000} 
                & \valfont{0.0206} 
                & \valfont{0.6931}                \\ \bottomrule

\multicolumn{6}{c}{} \\[0.4em] 
\rowcolor{gray!20} 
\multicolumn{6}{c}{\textbf{GRUT Top 5 prediction at 2014-01-01 ($\lambda=0.5$)}} \\ \midrule

\textbf{Ranking} & 1 & 2 & 3 & 4 & 5 \\ \midrule

\textbf{Image} &
\cellimg{Figures/CaseStudy/Rainbow.png} &
\cellimg{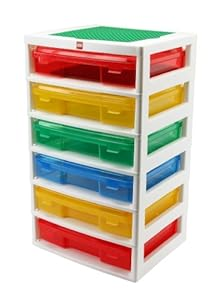} &
\cellimg{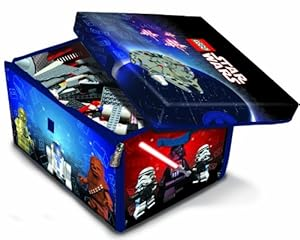} &
\cellimg{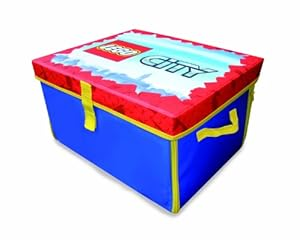} &
\cellimg{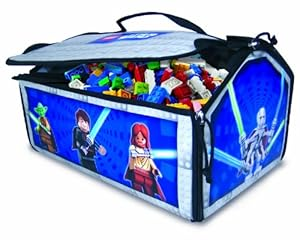} \\ \midrule

\textbf{Name}     & \valfont{\shortstack{Rainbow\\Loom}} 
                 & \valfont{\shortstack{LEGO 6-Case\\Storage Unit}} 
                 & \valfont{\shortstack{Star Wars\\Box}} 
                 & \valfont{\shortstack{LEGO City\\Box}} 
                 & \valfont{\shortstack{Star Wars\\Battle Bridge}} \\ \midrule

\textbf{Category} & \valfont{\shortstack{Toys \&\\Games}} 
                 & \valfont{\shortstack{Building\\Toys}} 
                 & \valfont{\shortstack{Vehicle\\Playsets}} 
                 & \valfont{\shortstack{Vehicle\\Playsets}} 
                 & \valfont{\shortstack{Toys \&\\Games}} \\ \midrule

\textbf{\(s_{\text{trend}}\)} 
                & \valfont{0.6931}
                & \valfont{0.1361} 
                & \valfont{0.0206} 
                & \valfont{0.0000} 
                & \valfont{0.0206}                 \\ \bottomrule
\end{tabular}

\caption{\ours’s top-5 predictions on the Toys dataset with and without trend-aware inference. The target item is marked with a red dotted line.}\label{fig:case_study_varying_lambda_case3}
\vspace{-3mm}
\end{table}

\subsection{Effect of Trend-aware Inference}\label{sec:app_case_trend}
Table~\ref{fig:case_study_varying_lambda_case3} illustrates how \ours~benefits from the trend score $s_\text{trend}$ to better capture user preference. The user had recently purchased the `\textit{LEGO Sorting Systems}', so various toy-related products appear as top recommendations when $\lambda$ = 0 in Eq.~\eqref{eq:final_score}. Considering temporal trends during inference, the ranking of trending items 
`\textit{Rainbow Loom}' was elevated, resulting in recommendations that closely aligned with the user preferences. This demonstrates that Trend-aware Inference enables the model to combine time-sensitive trends with the user's intrinsic preference, producing more accurate and timely recommendations.  Furthermore, the ability to control the influence of the $s_\text{trend}$ based on user needs highlights the practical advantage of the proposed method in terms of \textit{controllability}.

\end{document}